\documentclass[a4paper,12pt]{article}

\usepackage{amsmath,amsthm,amssymb,amsfonts}
\usepackage[margin=2.5cm]{geometry} 
\usepackage{graphicx}
\usepackage{caption}
\usepackage{subcaption}
\graphicspath{{img/}}
\usepackage[comma,authoryear]{natbib}

\usepackage{url}
\usepackage[table]{xcolor}
\usepackage{booktabs}
\usepackage{tabularx}
\usepackage{pdflscape}
\usepackage{placeins}
\usepackage[small]{titlesec}
\usepackage{bbm}
\usepackage{verbatim}
\usepackage{setspace}
\usepackage[bottom]{footmisc}
\usepackage{float}

\usepackage{mathrsfs}  
\usepackage{enumitem}
\usepackage{eurosym}
\usepackage{xr-hyper}

\def\sym#1{\ifmmode^{#1}\else\(^{#1}\)\fi}

\floatstyle{plaintop}
\restylefloat{table}

\usepackage{array,multirow}

\usepackage{longtable}	
\usepackage{lscape}

\usepackage[autostyle=false, style=english]{csquotes}

\usepackage[hidelinks]{hyperref}

\MakeOuterQuote{"}

\title{The Spoils of Algorithmic Collusion:\\ Profit Allocation Among Asymmetric Firms}
\author{Simon Martin\thanks{simon.martin@univie.ac.at. University of Vienna, Department of Economics, also affiliated with CESifo and CEPR} \and Hans-Theo Normann\thanks{normann@hhu.de. Düsseldorf Institute for Competition Economics (DICE), Heinrich-Heine-Universität Düsseldorf and Max Planck Institute for Research on Collective Goods, Bonn} \and Paul Püplichhuisen\thanks{pueplichhuisen@dice.hhu.de. Düsseldorf Institute for Competition Economics (DICE), Heinrich-Heine-Universität Düsseldorf, also affiliated with E.CA Economics GmbH}\and Tobias Werner\thanks{werner@mpib-berlin.mpg.de. Center for Humans and Machines at the Max Planck Institute for Human Development, Berlin\\\indent Computational infrastructure and support were provided by the Centre for Information and Media Technology at Heinrich Heine University Düsseldorf.}}
\begin{document}

\date{\today}

\maketitle

\begin{abstract}\noindent We study the propensity of independent algorithms to collude in repeated Cournot duopoly games. Specifically, we investigate the predictive power of different oligopoly and bargaining solutions regarding the effect of asymmetry between firms. We find that both consumers and firms can benefit from asymmetry. Algorithms produce more competitive outcomes when firms are symmetric, but less when they are very asymmetric. Although the static Nash equilibrium underestimates the effect on total quantity and overestimates the effect on profits, it delivers surprisingly accurate predictions in terms of total welfare. The best description of our results is provided by the equal relative gains solution. In particular, we find algorithms to agree on profits that are on or close to the Pareto frontier for all degrees of asymmetry. Our results suggest that the common belief that symmetric industries are more prone to collusion may no longer hold when algorithms increasingly drive managerial decisions.
\end{abstract}

\bigskip\bigskip



\onehalfspacing
\pagebreak

\section{Introduction}
\label{sec:Introduction}

It is well known that firms find it in their interest to (tacitly) collude, but that asymmetries are an obstacle to the successful implementation of collusive schemes. While firms may suppress output or set higher prices to achieve higher profits than those achieved under competition, they often differ along several dimensions, such as their cost structure, capacities, logistics, production networks, sales force, or product quality. Indeed, asymmetries seem to be the rule rather than the exception\footnote{
    For collusion studies on this point, see \cite{grout2005predicting,levenstein2006determines}, and \cite{davies2011tacit}.}, 
and the conventional wisdom is that such asymmetries make (tacit) collusion more difficult \citep{ivaldi2003economics}.  

The challenge of collusion in asymmetric industries is two-dimensional. On the one hand, firms must (tacitly) agree on total output, and on the other hand, they must decide how to share the spoils of such collusion. In the case of a symmetric industry, collusion is easier because firms have the same preferences regarding total output, and sharing profits equally seems an intuitive and focal solution to the bargaining problem. In contrast, when firms are asymmetric, they disagree about the collusive level of total output, and profit sharing may prove cumbersome: Joint payoff maximization may require the inefficient firms to shut down. However, if all firms produce positive quantities, the outcome is inefficient, and a convex bargaining frontier emerges \citep{Bishop1960,Schmalensee1987,Tirole1988}. Coordinating on the two dimensions without explicit communication is usually considered to be difficult.\footnote{In oligopoly models with asymmetric capacities, firms usually do not disagree on total output (or price), but the bargaining problem persists. With differentiated price competition, joint payoff maximization would require side payments, but the inefficient firms do not necessarily have to shut down. }  

The existing literature provides some insights into how asymmetric firms might collude, but there are limits to how the bargaining problem can be addressed.  
The theoretical literature mainly answers how asymmetric firms collude \textit{optimally} \citep{harrington1991determination, miklos2011optimal}, with or without side payments. However, the conditions under which asymmetric firms may be expected to find these optimal solutions to the bargaining problem, or to reach any alternative bargaining outcome, remain largely unexplored. The empirical literature on how firms maintain collusion in asymmetric industries, and in particular how they divide the spoils of collusion without side payments\footnote{\cite{clark2013collusion} find that the Canadian gasoline cartel implemented transfers between firms by adjusting delays during price changes.} suffers from the availability of appropriate data. Assessing the extent of tacit collusion is difficult without robust counterfactuals, and cartel studies may be biased because they are based on detected antitrust cases. Several laboratory experiments \citep{fonseca2008mergers,argenton2012collusion,harrington2016relative,fischer2019collusion} have addressed this issue. A key finding of these experiments is that tacit collusion fails in asymmetric settings, but they cannot explain what successful collusion might look like. 

In this paper, we propose a novel approach to this gap in the literature. Specifically, we analyze how self-learning algorithms share the profits of potentially colluding asymmetric firms. Methodologically, we solve the two-dimensional collusion problem by conducting a series of simulations with Q-learning algorithms \citep{watkins1989learning, watkins1992q}.

How an algorithm solves the simultaneous coordination and bargaining problems among asymmetric firms, is a subject of growing importance. Firms' decisions are increasingly being outsourced to algorithms across a variety of industries.\footnote{For instance, similar self-learning algorithms to the ones we consider are used in pricing \citep{liu2019dynamic,chen2023reinforcement}, ride sharing optimization \citep{qin2022reinforcement}, or inventory management in e-commerce \citep{madeka2022deep}.} Furthermore, even in situations where human decision-makers remain in control, they frequently receive algorithmic recommendations \citep[see, for instance,][]{garcia2022demand, huelden2024human}. 

We run several different parameterizations of the repeated asymmetric Cournot duopoly. Starting from a symmetric baseline variant, we explore six different degrees of asymmetry between firms, and we consider two setups that differ in their parameterization. The main parameterization setup explores increasingly asymmetric firms while keeping the total output of the (static) Cournot-Nash equilibrium constant. In the second parameterization setup, the marginal cost parameter of the efficient firm is kept constant, so that monopoly output remains invariant in this setting. We run Q-learning simulations for these fourteen variants for different exploration and learning parameters, and for both algorithms with and without memory. 

Our main findings are as follows. Somewhat surprisingly, both consumers and firms can benefit from asymmetry. As firms become more asymmetric, the more efficient firm produces additional output. Some of these efficiency gains are indirectly shared with consumers. However, the algorithms in our setting do not implement fully efficient outcomes, according to which the inefficient firm is entirely inactive. Relative to joint profit maximization, algorithms produce more competitive outcomes when firms are symmetric, but less when they are asymmetric. Intuitively, the extreme benchmark of one firm staying inactive would require some form of compensation, for example, through side payments, which cannot be captured in our environment. 

On the other extreme, Nash equilibrium (which assumes absence of coordination) underestimates the effect of asymmetry on total quantity but overestimates the effect on profits. Nevertheless, the two effects are similar in magnitude, and hence Nash equilibrium delivers surprisingly accurate predictions in terms of total welfare. 

Equal relative gains best describe our results, a solution criterion by \cite{roth1979game}. The idea is that firms profit from collusion proportionally to their respective disagreement profits, that is, the profits they would obtain under competition. Intuitively, the firm that expects to make the largest profits under competition should experience a proportional increase in profits under collusion. Although the Q-learning agents in our simulation tend to produce more, on average, than predicted by equal relative gains, this appears to be primarily a level effect. In particular, the comparative statics of increasing asymmetry are well captured by equal relative gains. 

Additionally, our results show that the profit allocation for all degrees of asymmetry lies on the Pareto frontier. In contrast, the Nash equilibrium does not. We interpret this as an indication that the outcome that algorithms converge to is well explained by bargaining solutions such as equal relative gains.

Our results have important implications for competition policy and the regulation of digital markets. Namely, our findings suggest that the common belief that symmetric industries are more prone to collusion may no longer hold when algorithms increasingly drive managerial decisions \citep{leisten2021algorithmic,garcia2022demand,hunold2023algorithmic}.

On top of providing us with the possibility of addressing long-standing questions regarding asymmetry in collusion, this setup also allows us to address current debates about algorithmic collusion   \citep{ezrachi2017artificial,harrington2018developing,harrington2022effect, OECD2023italy, OECD2023algorithms} as supra-competitive prices are detrimental to consumers and economic welfare.\footnote{
    Pricing algorithms can also have a positive impact on economic welfare: Algorithms are well equipped to handle the vast amount of data available online about competitors and customers. They can adjust prices dynamically, support consistent pricing strategies, and respond immediately to changes in the market environment \citep{OECD2023algorithms}. These efficiencies could ultimately benefit consumers. }
Supporting the concerns, recent research by \cite{calvano2020artificial} and \cite{klein2021autonomous} reveals that self-learning algorithms can learn to collude in repeated pricing games. This not only concerns supra-competitive prices, but algorithms can also devise strategies that use punishments to discourage deviations from agreed-upon behavior. So, price algorithms may tacitly collude. This makes this a challenge for competition policy because these self-learning algorithms were never instructed to do so. Their work has been extended by \cite{hettich2016algorithmic} and \cite{johnson2023platform}, among others. Notably, most of these papers study price competition (with a few exceptions, as elaborated on below).  

The remainder of this paper is structured as follows. We provide an overview of how our work relates to other literature in the next section. We explain our model and experimental setup in Section \ref{sec:Model} and Section \ref{sec:experiments}, respectively. Our main results, mechanism, and several robustness checks are presented in Section \ref{sec:results}. Section \ref{sec:Conclusion} concludes. 

\section{Related Literature} 

Relative to the existing theoretical literature on collusion of asymmetric firms \citep{harrington1991determination, miklos2011optimal}, our work offers novel insights into how the bargaining problem of asymmetric firms is resolved. We show that outcomes tend to be well explained by the concept of equal relative gains, in particular, the effect of increasing asymmetry. Similarly, the experimental literature \citep{fonseca2008mergers, argenton2012collusion, harrington2016relative, fischer2019collusion} typically only contrast symmetric with asymmetric settings, but does not vary the degree of asymmetry as pronounced as we are doing. We thus offer novel insights into the comparative statics of different degrees of asymmetry.

We also contribute to the literature on collusion of self-learning algorithms. \citet{calvano2020artificial} analyze Q-learning in markets with logistic demand, horizontal product differentiation, and simultaneous price competition. For a variation with asymmetric costs, the results imply that relative to the competitive Nash equilibrium, the less efficient firm gains more compared to the more efficient firm under price competition. \citet{klein2021autonomous} consider sequential price competition with symmetric Q-learning agents. They show that algorithms can learn to play supra-competitive price and punishment strategies that can rationalize the prices as a subgame perfect Nash equilibrium. \citet{waltman2008q} consider competition in quantities but focus on memoryless algorithms, which limits their ability to learn collusive strategies. Similarly, \citet{abada2023artificial} analyze repeated Cournot games with symmetric costs for electricity storages using Q-learning algorithms. They find that algorithms were able to reach supra-competitive profits above the three-player Cournot equilibrium. While they consider agents that are asymmetric in terms of their storage capacity as a robustness check, they do not vary the degree of asymmetry in a systematic way. Also, \citet{wang2022will} and \citet{taywade2022using} analyze, among other specifications, Cournot competition in duopolies with cost asymmetries, but without considering the degree of asymmetry. In our paper, we explicitly analyze how variations in the degree of asymmetry yield different market outcomes. Furthermore, we benchmark different theoretical predictions against each other to test which theory best predicts collusion among algorithms in this environment.

There is also increasing evidence that algorithmic adoption is linked to anti-competitive behavior from empirical studies. For instance, \citet{assad2024algorithmic} find that an increase in the adoption of pricing algorithms is linked to an increase in mark-ups in the German gasoline markets. Other empirical studies focus on e-commerce platforms \citep{wieting2021algorithms, musolff2022algorithmic} and similarly find evidence of anti-competitive effects from algorithms. By focusing on a controlled environment in simulations, we can analyze the mechanisms behind those phenomena when firms are asymmetric in a structured way.

While there are several experimental studies that either compare algorithmic and human collusion \citep{kasberger2023algorithmic, werner2022algorithmic} or study the interaction between algorithms and humans \citep{normann2021hybrid, schauer2022competition}, they do not consider Cournot competition.

Also in experimental studies without algorithms, the focus is often not on asymmetries among the firms. The closest related study is \citet{fischer2019collusion}, who conduct an experiment on Cournot competition with cost asymmetries. They find that participants could not sustain collusion in this asymmetric environment without communication, while supra-competitive outcomes with a relative advantage for the less efficient firm could be achieved in the case of communication. In our paper, we follow a similar approach to \citet{fischer2019collusion} and analyze how firms can sustain collusion, but we focus on Q-learning algorithms instead.

\section{Model}\label{sec:Model}

We consider a standard Cournot duopoly, allowing for asymmetries in marginal costs. There are two firms $i\in \{L, H\}$ which simultaneously choose non-negative quantities $q_L$ and $q_H$ from the interval $[0, q^{max}]$. The firms only differ by their constant marginal production costs $c_i$, with $c_H \ge c_L \ge 0$. The products are homogeneous, and the inverse demand function is given by 

\begin{equation*}
    p(Q)= \max \left\{a-bQ, 0 \right\}, 
\end{equation*}
where $a$ and $b$ denote the intercept and slope parameters, respectively, and $Q = q_H + q_L$ denotes total quantity. Firm $i$'s profit is given by 
\begin{equation}
    \pi_i = p(Q) q_i - c_i q_i.
    \label{eq:profit}
\end{equation}
We will analyze producer surplus (PS), consumer surplus (CS), and total surplus (TS):
\begin{align*}
    PS & = \pi_H + \pi_L\\
    CS & = \frac{a - p(Q)}{2} Q = \frac{b Q^2}{2} \\
    TS & = PS + CS
\end{align*}

Given this model, we consider possible market outcomes under different modes of competition in the following. These will serve as benchmarks, when discussing the results in Section \ref{sec:results}.

\paragraph{Nash equilibrium.} Taking first-order conditions in the profit expression \eqref{eq:profit} immediately yields the standard static unique Nash equilibrium of the game:
\begin{align*}    
    q_i^{NE} = \frac{a-2c_i+ c_{j}}{3b}
\end{align*}
and equilibrium profits are $\pi^{NE}_i = b(q_i^{NE})^2$.

\paragraph{Joint Profit Maximization.} For joint profit maximization, only the cost-efficient firm $L$ is active and produces its monopoly quantity, whereas the inefficient firm $H$ stays out of the market. So, $i=L$ produces the monopoly quantity
\begin{equation}
    q_i^{M}=\frac{a-c_i}{2b}
    \label{eq:monoopoly-quantity}
\end{equation}
resulting in total industry profits $\Pi^M = \pi^M_L + 0 =b(q_L^M)^2$.

\paragraph{Alternating monopoly.} For this benchmark, we assume that one firm becomes the monopolist with probability $\omega$, and the other firm stays out of the market altogether. In our application, we assume that $\omega=0.5$, that is, each of the firms is a monopolist with probability $50\%$, chooses $q_i^M$ and earns $\pi^M_i$. For example, they may coordinate on being the monopolist alternately. 

\bigskip

The previous three solutions are standard oligopoly benchmarks, but they may have infeasible profit allocation implications for colluding firms. The following bargaining solutions represent alternative predictions for the level and distribution of profits. Later in the paper, we sometimes refer to ``oligopoly benchmarks'' and ``bargaining solutions'' to distinguish the two sets of solutions.

Bargaining solutions assume that firms coordinate on the Pareto frontier. Each point on the Pareto frontier maximizes the profit that can be earned by one firm given a profit target for the other firm. To obtain the Pareto frontier, we rewrite the profit function as $q_i=\pi_i/(p-c_i)$, and then sum over both firms: 
\begin{equation}
    Q = \frac{\pi_i}{p-c_i}+\frac{\pi_j}{p-c_j}.
\end{equation}
Solving for $\pi_j$ with $Q=a-p$ provides the Pareto frontier
\begin{equation}
    \overline{\pi}_j(\pi_i) = \max_p \left(a-p-\frac{\pi_i}{p-c_i}\right)(p-c_j).
      \label{eq:Pareto}
\end{equation}
That is, for any profit of firm $i$, a market price $p$ can be found that maximizes the profit of firm $j$.
Since costs are asymmetric and both firms produce positive quantities, the Pareto frontier is convex \citep{Bishop1960,Schmalensee1987,Tirole1988}. Although all bargaining concepts below assume that the solution lies on the Pareto frontier, they differ in the choice of a point on this frontier. 

\paragraph{Kalai-Smorodinsky.} \cite{kalai1975other} assume a profit split relative to the maximum profit a firm can achieve ($\pi_i^M$) and the profit under disagreement ($\pi_i^{dis}$), that is:
\begin{align}
    \frac{\pi_i^{KS}-\pi_i^{dis}}{\pi_i^M-\pi_i^{dis}} &= \frac{\pi_j^{KS}-\pi_i^{dis}}{\pi_j^M-\pi_i^{dis}}.
    \label{eq:KS}
\end{align}
In this specification, the maximum profit achievable is the one-firm monopoly profit.
As disagreement profits ($\pi_i^{dis}$), we consider (i) non-cooperative Nash profits ($\pi_i^{NE}$) and (ii) min-max disagreement profits ($\underline{\pi_i}$).
The latter are defined as the maximum profit a firm can achieve given that the other firm produces the maximum quantity at the upper bound of the action space, $q^{max}$: 
\begin{equation*}
    \underline{\pi_i}=\max_{q_i} \ \pi_i(q_i, q^{max}).
\end{equation*}
The Kalai-Smorodinsky solution to \eqref{eq:KS} is then implicitly defined by plugging \eqref{eq:KS} into the Pareto frontier expression in \eqref{eq:Pareto}, and maximizing with respect to $p$. 

\paragraph{Equal relative gains.} This bargaining concept of \cite{roth1979game} also follows the idea of profit sharing relative to the point of disagreement. However, it ignores the maximum (monopoly) profits: 
\begin{equation*}
    \frac{\pi_i^{ERG}}{\pi_i^{dis}} = \frac{\pi_j^{ERG}}{\pi_j^{dis}}.
\end{equation*}
As with the Kalai-Smorodinsky solution, we consider here as disagreement profits (i) Nash profits and (ii) min-max profits.

\paragraph{Equal split.} An equal split solution according to \cite{roth1979game} implies that the bargaining point on the Pareto frontier ensures both firms receive the same payoff: 
\begin{equation*}
    \pi_i^{ES} =  \pi_j^{ES}.
\end{equation*}

\section{Experimental setup}
\label{sec:experiments}

\subsection{Main parameterization of the Cournot model}

We study seven different parameterizations, starting with symmetric firms and increasing the level of cost asymmetry between the two firms. Our main parameterization is shown in Table \ref{tab:Benchmark}.\footnote{As a robustness check, we also examine an alternative parameterization setup in Section \ref{sec:robustness-checks}, where the marginal cost of the efficient firm is held constant, ensuring that monopoly output stays unchanged as we increase the degree of asymmetry.} Throughout, we use an intercept $a= 91$ and a slope $b=1$, which nests the cases studied in \cite{fischer2019collusion}. We start with a symmetric specification (Sym) in which both firms have marginal costs of $c = 19$, resulting in Nash quantities of $q_L^{NE} = q_H^{NE} = 24$ and hence total Nash output of $Q^{NE} = 48$. Under joint profit maximization, output would be limited to $Q^M = 36$. 

In specifications Asym1 to Asym6, we gradually make firms more asymmetric by decreasing $c_L$ and simultaneously increasing $c_H$ so that total Nash quantities $Q^{NE}$ remain unaffected. However, under Nash competition, firms' output quantities also become increasingly asymmetric. For example, in Asym6 almost all total output is produced by the more efficient firm. Furthermore, total monopoly quantities $Q^M$ increase as firms become more asymmetric, since the only active firm $L$ has decreasingly lower marginal costs.

\begin{table}[ht]
\caption{Cost parameterization.}
\label{tab:Benchmark} 
\begin{longtable}{>{\raggedright\arraybackslash}p{0.15\textwidth}  l c c c c c c c} 
        \hline 
    & Sym & Asym1 & Asym2 & Asym3 & Asym4 & Asym5 & Asym6 \\
    \hline 
    $c_L$ & 19 & 16 & 13 & 10 & 7 & 4 & 1 \\
    $c_H$ & 19 & 22 & 25 & 28 & 31 & 34 & 37\\
    $q_L^{NE}$ & 24 & 27 & 30 & 33 & 36 & 39 & 42  \\
    $q_H^{NE}$ & 24 & 21 & 18 & 15 & 12 & 9  & 6 \\
    $Q^{NE}$ & 48 & 48 & 48 & 48 & 48 & 48 & 48 \\
    $Q^M$ & 36 & 37.5 & 39 & 40.5 & 42 & 43.5 & 45 \\
    \hline
\end{longtable} 
\end{table}

For each of our specifications, the solutions discussed in Section \ref{sec:Model} have different predictions. We will investigate outcomes based on experiments with Q-learning algorithms.

\subsection{Q-Learning}
\label{sec:Qlearning-short}

Similar to \citet{calvano2020artificial} and others, we utilize Q-learning algorithms. Each period, the Q-learning agents can choose from a finite and discrete action space $a_{i,t}\in A$.
Further, $s_{i,t} \ \in S$ denotes the state of the environment in the period $t$ as an element of the set of possible states. We assume that an agent $i$ has bounded knowledge of the actions chosen by all other agents over the past $k$ periods. Then a state $s_{i,t} = \{\mathbf{a}_{t-1}, \dots, \mathbf{a}_{t-k} \}$ contains the actions of all agents for the past $k$ periods, where $\mathbf{a}_t$ is a vector of the actions of all agents in period $t$.

Each agent aims to maximize the discounted expected future profits over the action set $A$, given the current state $s_{i,t}$. For a given period, the reward from picking a certain action in a certain state is denoted by $\pi_{i,t}$.
For Q-learning algorithms, we can write
\begin{equation}
    Q_{i,t}(s_{i,t}, a_{i,t}) = \mathbb{E}[\pi_{i,t}|s_{i,t}, a_{i,t}] + \delta\mathbb{E}[\max_{a} Q(s_{i,t+1}, a)|s_{i,t}, a_{i,t}]
\end{equation}
Since the set of states and the set of actions are finite in our setup, the Q-function is a $|S| \times |A|$ matrix in which $Q_{i,t}(s_{i,t}, a_{i,t})$ is the current approximation of the stream of future discounted rewards associated with choosing action $a_{i,t}$ in state $s_{i,t}$.

When choosing action $a_{i,t}$ in state $s_{i,t}$ and receiving a reward $\pi_{i,t}$, the Q-matrix then gets updated as follows:
\begin{equation}
\label{eq:Qlearning}
    Q_{i,t+1}(s_{i,t}, a_{i,t}) = (1 - \alpha) Q_{i,t}(s_{i,t}, a_{i,t}) + \alpha \left( \pi_{i,t} + \delta \max_{a} Q_{i,t}(s_{i,t+1}, a) \right)
\end{equation}
where $\alpha \in (0,1)$ denotes the learning parameter. It governs how much weight the algorithm gives to the information it already has about the environment compared to newly arriving information in the current round.

When choosing an action $a_t$, an agent faces a trade-off between relying on past experiences (\textit{exploitation}) versus exploring the action field for each state to find the optimal action-state combination (\textit{exploration}). We incorporate experimentation in an $\epsilon$-greedy process. That is, an agent chooses the current optimal action $a^*_t$ for state $s_t$ with probability $(1-\epsilon_t)$. An agent chooses a random action with probability $\epsilon_t = e^{-\beta t}$, where $0 < \beta < 1$ is the exploration decay.

\subsection{Methodology}
\label{sec:methodology}
We implement our study by having two independent Q-learning agents play a repeated game (with cost parameterization from Table \ref{tab:Benchmark}) against each other. The same approach is taken, for example, in \cite{calvano2020artificial}, \cite{klein2021autonomous}, \cite{johnson2023platform} and \cite{kasberger2023algorithmic}.

Both agents use the same ``technology'', that is, a learning rate $\alpha$, experimentation with exponential decay at rate $\beta$, and a common discount factor $\delta = 0.95$. In our baseline specification with \textit{moderate exploration}, we let $\alpha = 0.15$ and $\beta = 3.41\times 10^{-6}$ as in \cite{calvano2019algorithmic}.\footnote{We chose a baseline $\beta$ that provides the same expected times a cell would be visited by pure random exploration, that is, $\nu = ((m-1)^n) / (m^{kn+n+1} \cdot (1 - e^{-\beta(n+1)}))\approx21$, which yields $\beta = 3.41\times 10^{-6}$. A slight adjustment is needed relative to \cite{calvano2019algorithmic} as our experimental setup involves $m=16$ choices instead of $m=15$ choices. We provide several robustness checks in these dimensions in Section \ref{sec:robustness-checks}.}

Both agents may condition their actions on the previous round's actions, that is, the memory length $k=1$. We discretize the action space in steps of three, that is, $q_i \in [0,3,6, \ldots, 45]$, so $m=16$, and we initialize the Q-matrix with random values from the uniform distribution on $[0,1\times10^{-7}]$. We run the simulation until convergence, defined as not changing the best action for each state $s \in S$ over 100,000 subsequent periods. Upon convergence, we let the agents play 1,000 rounds. For each specification and parameterization, we repeat 1,000 simulations, and then consider the average across those simulations runs.

Given the results we obtain from our simulations, we compute a range of distance measures (goodness of fit) for several key outcome variables. Specifically, we are interested in the following four outcome variables $y \in \{Q, PS, CS, TS\}$, where $Q$ denotes total quantity, $CS$ the respective consumer surplus, $PS = \pi_L + \pi_H$ the producer surplus, and $TS = CS + PS$ the total surplus.

For each of these outcome variables, we consider the seven specifications $\sigma  \in \Sigma = {\{sym, asym1, asym2, asym3, asym4, asym5, asym6\}}$. For each $y$ and $\sigma$, there is a simulation outcome, $y_\sigma^s$, as well as eight theoretical bargaining solutions $y_\sigma^b$, for $b \in \{n, m, am, erg, \allowbreak es, ks, erg_n, ks_n\}$, denoting (static) Nash, monopoly, alternating monopoly, equal relative gains, equal split, Kalai-Smorodinsky, equal relative gains with Nash disagreement profit and Kalai-Smorodinsky with Nash disagreement profit, respectively.

For each $y$ and $b$, we compute both an \textit{average squared distance} to the simulation result (capturing the overall goodness of fit), given by
\begin{align*}
    \overline{y^b} = \frac{1}{|\Sigma|} \sum_{\sigma \in \Sigma} \left(y_\sigma^s - y_\sigma^b \right)^2
\end{align*}
as well as a \textit{average squared normalized distance} to the simulation results (capturing the goodness of fit of the comparative statics), given by
\begin{align*}
    \widehat{y^b} = \frac{1}{|\Sigma|} \sum_{\sigma \in \Sigma} \left(\widehat{y_\sigma^s} - \widehat{y_\sigma^b} \right)^2
\end{align*}
where $\widehat{y_\sigma^b}$ denotes normalized outcome variables to the symmetric case:
\begin{align*}
\widehat{y_\sigma^b} =  \frac{y_\sigma^b}{y_{sym}^b}. 
\end{align*}

\section{Results}
\label{sec:results}

\subsection{Simulation results}

This section provides an overview of our simulation results, focusing on the collusive potential of algorithms. We also examine the respective welfare implications. 

\begin{figure}[hbt!]
\begin{center}
\begin{subfigure}{.5 \textwidth}
  \centering
  \includegraphics[width=0.95\linewidth]{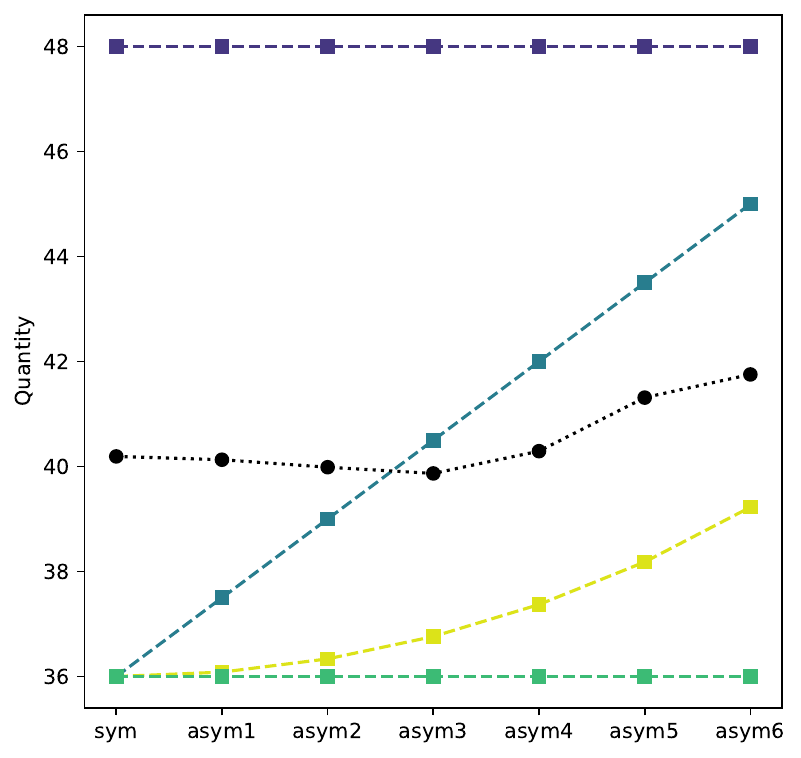}
  \caption{Total Quantities}
  \label{fig:level:q}
\end{subfigure}%
\begin{subfigure}{.5 \textwidth}
  \centering
  \includegraphics[width=0.95\linewidth]{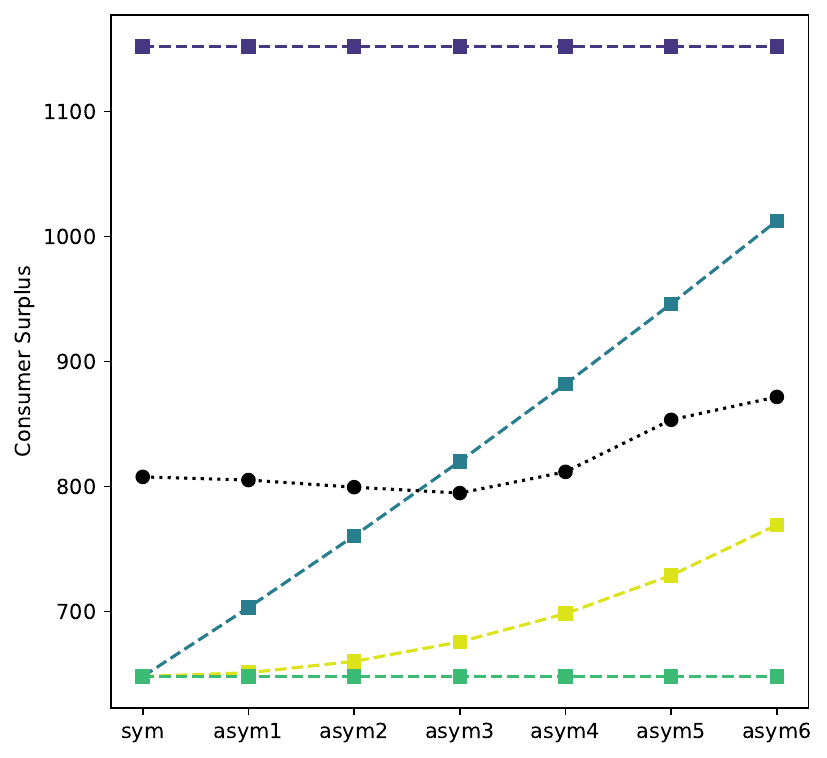}
  \caption{Consumer Surplus}
   \label{fig:level:cs}
\end{subfigure}
\begin{subfigure}{.5 \textwidth}
  \centering
  \includegraphics[width=0.95\linewidth]{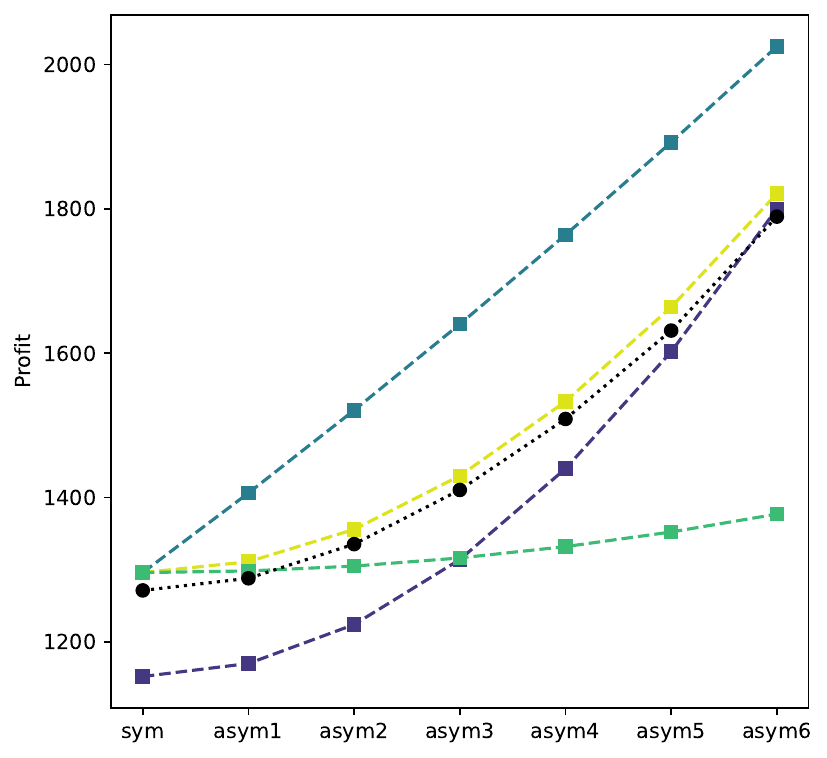}
  \caption{Total Profits}
  \label{fig:level:pi}
\end{subfigure}%
\begin{subfigure}{.5 \textwidth}
  \centering
  \includegraphics[width=0.95\linewidth]{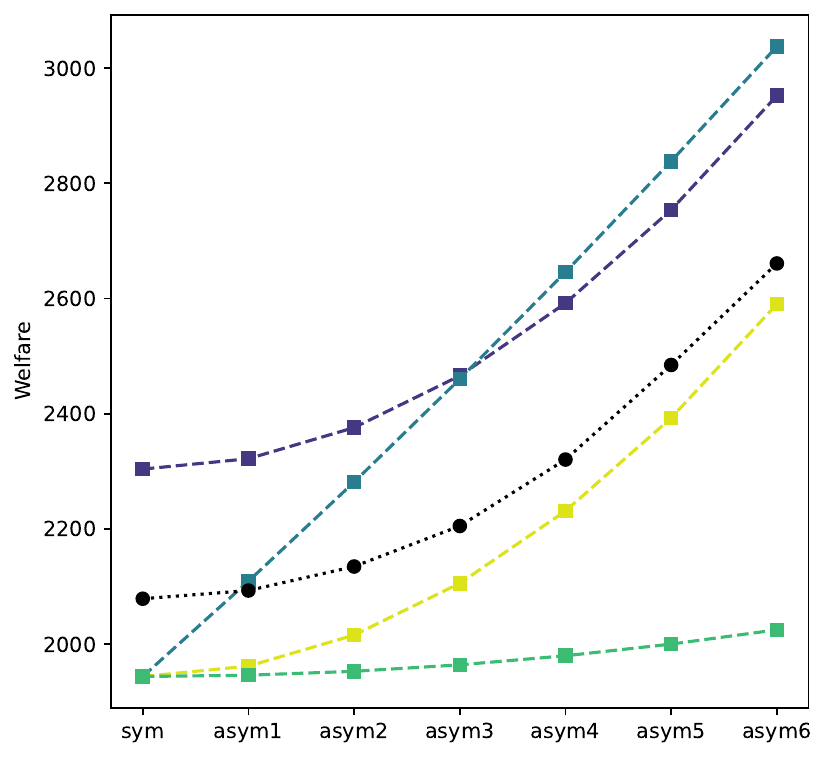}
  \caption{Total Welfare}
\end{subfigure}
\begin{subfigure}{.95 \textwidth}
  \centering
  \includegraphics[width=0.85\linewidth]{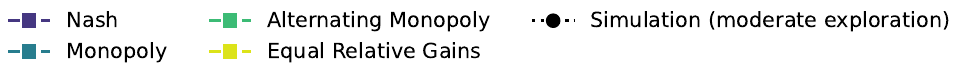}
\end{subfigure}
\caption{Simulation results for different degrees of asymmetry across different outcome variables. The figure compares the simulation results to four theoretical benchmarks: Nash equilibrium, monopoly, alternating monopoly, and equal relative gains. For comparisons with additional theoretical benchmarks, see Figure~\ref{fig:res:1_all} in Appendix \ref{sec:additional-figures}.}
\label{fig:res:1}
\end{center}
\end{figure}

\paragraph{Evidence of collusive outcomes.} As shown in Figure~\ref{fig:res:1}, panel (a), total quantities across asymmetry specifications center around $Q\approx 40$. This falls between the theoretical benchmarks of static Nash level (48) and the collusive benchmark of alternating monopoly. The output produced by algorithms is restricted relative to the competitive benchmark. It provides evidence that the outcomes obtained by algorithms are collusive

We next investigate the effect of increasing asymmetry between firms. According to our simulation results, total quantity \textit{increases} in asymmetry. However, for low levels of asymmetry, the simulated quantities are \textit{above} the joint profit-maximizing solution (monopoly quantities), that is, the algorithms are \textit{less collusive} than they could be. For high levels of asymmetry, in contrast, the simulated quantities are \textit{below} the monopoly benchmark. In other words, algorithms produce more competitive outcomes (than suggested by models of collusion) when firms are symmetric but less collusive outcomes when they are very asymmetric. Importantly, this implies that asymmetric markets are relatively more collusive than symmetric ones. 

While these results indicate that algorithms yield collusive outcomes, they do not explain the underlying mechanisms. We abstain from investigating these forces for the time being and continue analyzing outcomes for the moment. We explore these collusive mechanisms further in Section \ref{sec:mechansisms} 

\paragraph{Welfare Effects.} We next investigate the welfare effects of market asymmetry on the outcomes obtained by algorithms. In panel (b) of Figure \ref{fig:res:1}, we depict the consumer surplus. In the homogeneous goods market we are investigating, consumer surplus is proportional, so the previous statements regarding total quantity translate directly into the effect on consumer surplus. From a consumer's point of view, algorithms lead to outcomes that are not as favorable as those under (static Nash) competition, but better than those that would emerge from joint profit maximization. 

The comparison of total profits, shown in panel (c),  is less straightforward. In the presence of cost asymmetries, it not only matter which quantity is produced, but also which of the two firms produces which output. We find that total profits under simulated algorithms are smaller than under joint profit maximization (monopoly), but notably higher than those under competition. Thus, again, the outcome under simulated algorithms lies between the benchmarks of different modes of competition. For all cases, the monopoly solution constitutes the upper bound of what total profit could be achieved at maximum by the two firms as it implies only the low-cost firm provides positive quantities. The lower bound is the Nash solution for lower degrees of asymmetry and the alternating monopoly solution for higher degrees of asymmetry (\textit{asym4, asym5, asym6}).

For the symmetric case, the solutions are in the interval $[1152, 1296]$, where the lower bound corresponds to the Nash solution and the upper bound to the monopoly solution. With total profits of about 1275, our simulation results are close to the upper bound of the interval. In contrast, for the most asymmetric case, we observe total profits around 1800, which corresponds to the noncooperative Nash equilibrium. For all other cases in between, the algorithms achieve total profits well above the Nash equilibrium. In line with all theoretical benchmarks except equal split, total profits rise with the degree of asymmetry.

Given these results, the findings on total welfare, shown in panel (d), follow naturally. First, algorithms do not manage to appropriate welfare gains entirely. With strong cost asymmetry, total welfare would dictate the inefficient firm exiting the market. However, this outcome is unattainable without side payments, which are outside our model. Second, we find that total welfare can increase in cost asymmetry. Taken together, these results suggest that when algorithms take over managerial decisions, both consumers and firms can be better off in more asymmetric environments. 

\subsection{Explanatory power of oligopoly and bargaining concepts}
\label{sec:explanatory-power}
We now explore to what extent the various concepts suggested in the literature can explain the outcomes achieved by the algorithms. To this end, we investigate which solution concept best describes the results using the methodology introduced in Section \ref{sec:methodology}.

\paragraph{Level effects.} We first investigate level effects, that is, which solution concept is closest to our simulated results in absolute terms. We may consider the relative distance of each concept to the simulation outcomes in Figure \ref{fig:res:1}, separately for total quantities, consumer surplus, profits, and total welfare. Table \ref{tab:average_distances} shows a systematic evaluation by computing squared distances.

As asymmetry increases, bargaining solutions align increasingly well with simulated output, particularly equal relative gains (for both disagreement profits). Interestingly, the simulation results in terms of quantity are closest, on average, to those predicted by a monopoly (joint profit maximization). However, this averaging masks an important tendency: The monopoly prediction rises steeper in asymmetry than our simulation results. That is, the monopoly solution provides predictions significantly below the actual level for the symmetric and low asymmetry cases. However, its predictions rise well above the simulation results for higher degrees of asymmetry.

The equal split solution performs just as good or bad as the other bargaining solutions for the symmetric case. However, it performs increasingly worse for higher degrees of asymmetry. This highlights how an equal profit split might be a feasible solution when players are symmetric but is rather unlikely in the presence of (cost) asymmetries.

Total profits are predicted spot on by the equal relative gains solutions and Kalai-Smorodinsky with Nash disagreement profits for all degrees of asymmetry. Consequently, the average squared distance in Table~\ref{tab:average_distances} is small. In particular, they perform remarkably better than predicted by the oligopoly benchmarks. Equal relative gains and Kalai-Smorodinsky also provide relatively good predictions for total welfare.

Overall, concepts implicitly embedding a bargaining idea such as equal relative gains and Kalai-Smorodinsky provide a strong explanation for the outcomes generated by algorithms, despite underestimating total quantity. Similarly, while the static Nash equilibrium underestimates the impact of asymmetry on total quantity, it overestimates the effect on profits. These opposing biases nearly offset each other, resulting in relatively accurate predictions for total welfare

\begin{table}[htb!]
\centering
\caption{Average Squared Distances from Simulation Results (moderate exploration)}
\label{tab:average_distances}
\begin{tabular}{lrrrr}
\hline
 & Q & CS & $\Pi$ & W \\
\hline
Nash & 57 & 110,618 & 7,945 & 65,716 \\
Monopoly & 6 & 10,208 & 41,506 & 68,265 \\
Alternating Monopoly & 21 & 30,526 & 41,379 & 126,583 \\
Kalai-Smorodinsky & 34 & 48,309 & 40,517 & 163,873 \\
Equal Split & 58 & 76,802 & 145,750 & 415,075 \\
Equal Relative Gains & 12 & 17,384 & 654 & 11,558 \\
Kalai-Smorodinsky (Nash) & 12 & 17,134 & 823 & 11,355 \\
Equal Relative Gains (Nash) & 9 & 13,241 & 4,428 & 7,914 \\
\hline
\end{tabular}
\end{table}

\begin{figure}[hbt!]
\begin{center}
\begin{subfigure}{.5 \textwidth}
  \centering
  \includegraphics[width=0.95\linewidth]{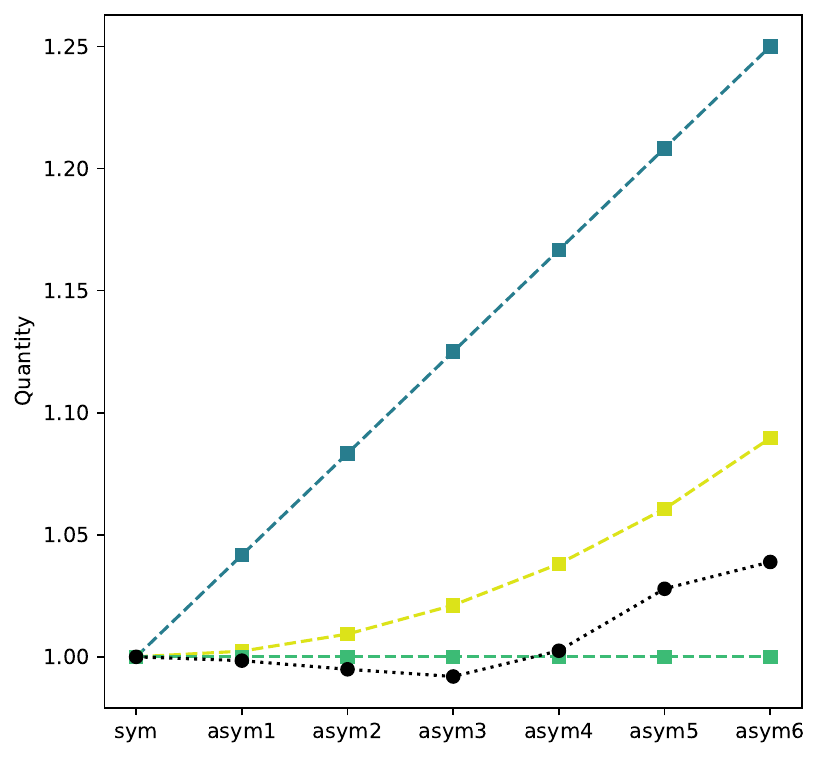}
  \caption{Total Quantities}
  \label{fig:comp:q}
\end{subfigure}%
\begin{subfigure}{.5 \textwidth}
  \centering
  \includegraphics[width=0.95\linewidth]{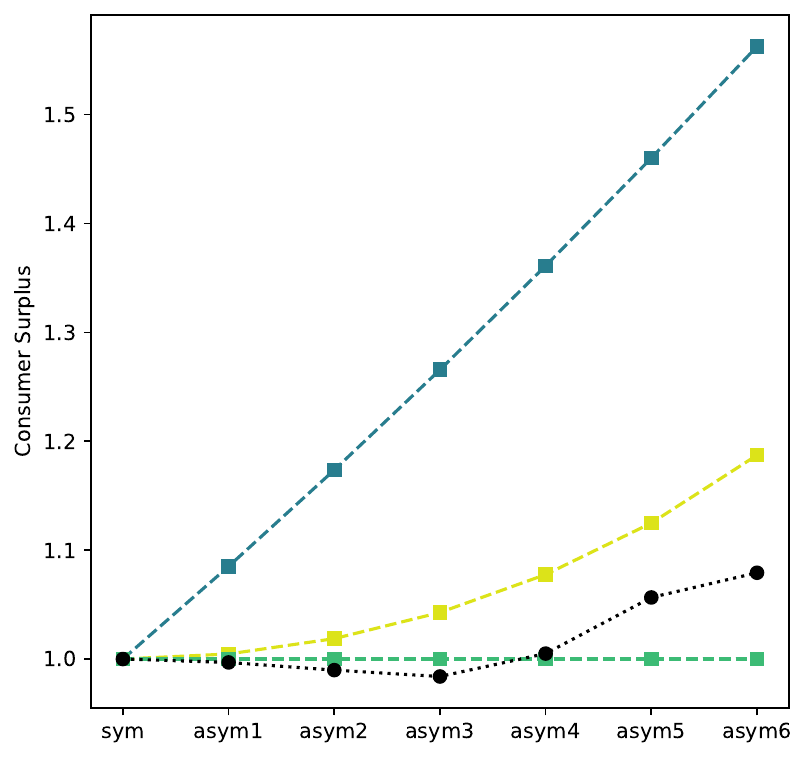}
  \caption{Consumer Surplus}
   \label{fig:comp:cs}
\end{subfigure}
\begin{subfigure}{.5 \textwidth}
  \centering
  \includegraphics[width=0.95\linewidth]{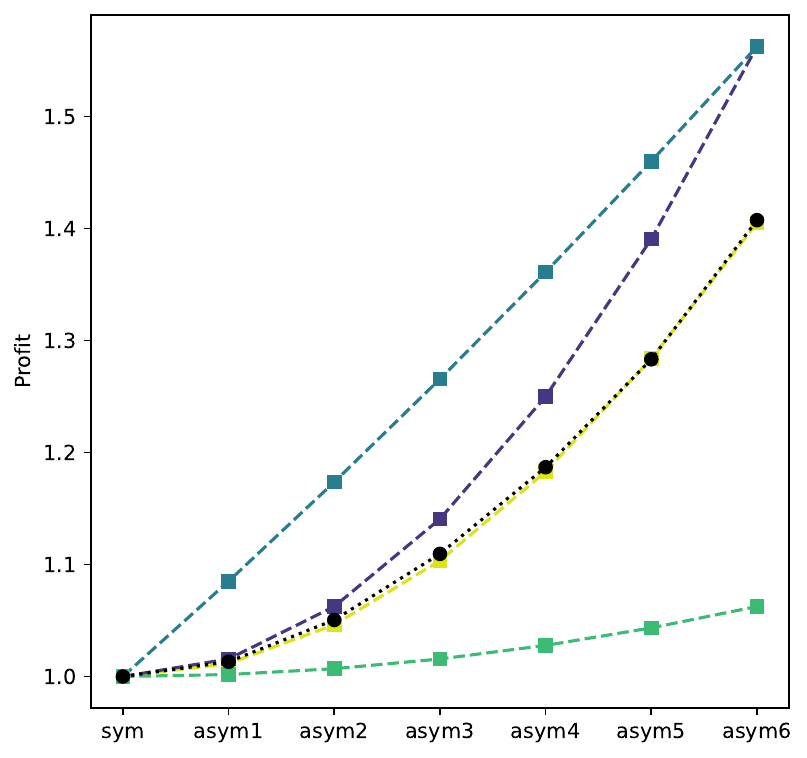}
  \caption{Total Profits}
  \label{fig:comp:pi}
\end{subfigure}%
\begin{subfigure}{.5 \textwidth}
  \centering
  \includegraphics[width=0.95\linewidth]{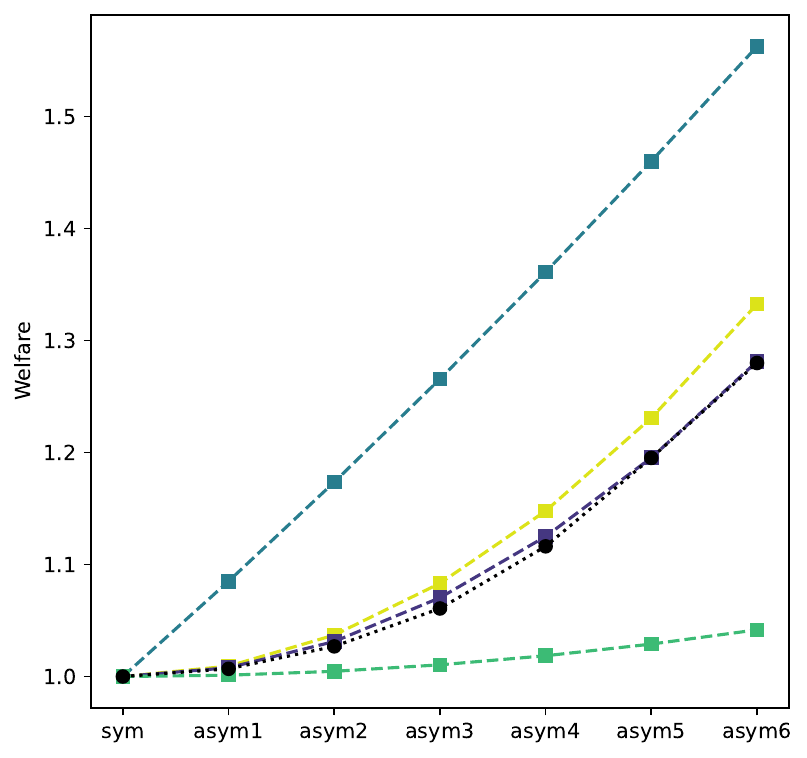}
  \caption{Total Welfare}
   \label{fig:comp:w}
\end{subfigure}
\begin{subfigure}{.95 \textwidth}
  \centering
  \includegraphics[width=0.85\linewidth]{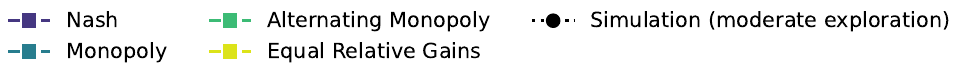}
\end{subfigure}
\caption{Simulation results (comparative statics; symmetric case = 1) for different degrees of asymmetry across various outcome variables. The figure compares the simulation results to four theoretical benchmarks: Nash equilibrium, monopoly, alternating monopoly, and equal relative gains. For comparisons with additional theoretical benchmarks, see Figure~\ref{fig:res:comp:full} in Appendix \ref{sec:additional-figures}.}
\label{fig:res:comp}
\end{center}
\end{figure}

\paragraph{Comparative statics.} We are particularly interested in the differential effect of increasing asymmetry on various market outcomes. Therefore, we now abstract from different levels across theories by considering outcomes relative to the benchmark of a symmetric specification. This allows us to isolate the comparative statics effect of changes in asymmetry. The results are shown in Figure \ref{fig:res:comp}, with a corresponding average squared distance calculation in Table \ref{tab:average_distances_normalized} in the Appendix.

Note that we designed the simulations so that the total quantities under Nash remain constant across all degrees of (a)symmetry, resulting in a horizontal line at value $1$ in Figure \ref{fig:res:comp}, panel (a), for the Nash equilibrium. The same is true for the alternating monopoly. Again, the trajectory of the simulation results is predicted best by equal relative gains and Kalai-Smorodinsky (with Nash deviation profits). Thus, these bargaining solutions do not only provide a best fit in terms of the overall (level) prediction accuracy but also a reasonably good fit for the relative change. 

For total profits, equal relative gains and Kalai-Smorodinsky perfectly capture the comparative statics of asymmetry perfectly. Similarly, the effect of asymmetry on total welfare is captured extremely well by the Nash prediction, as well as the bargaining concepts of equal relative gains and Kalai-Smorodinsky (with Nash). However, the good fit is masked by slight over- and underestimation of consumer surplus and total profits, respectively. In particular, the Nash prediction underestimates the effect on consumer surplus and overestimates the impact on total profits, while it is the opposite for the bargaining solutions. 

Taken together, we conclude that the best explanation for our results is provided by equal relative gains, and bargaining concepts more generally. Algorithms seem to ``agree'' on profits that are on or close to the Pareto frontier for all degrees of asymmetry.

\subsection{Mechanisms}
\label{sec:mechansisms}
A central problem for collusion of asymmetric firms concerns profit allocation. In contrast to collusion with symmetric firms where equal profit sharing is a natural candidate, profits of asymmetric firms under competition differ, making it unclear how profits should be split under collusion. This issue ultimately boils down to resolving an upfront bargaining problem over the spoils of collusion.

\begin{figure}[H]
\begin{center}
\begin{subfigure}{.5 \textwidth}
  \centering
  \includegraphics[width=0.95\linewidth]{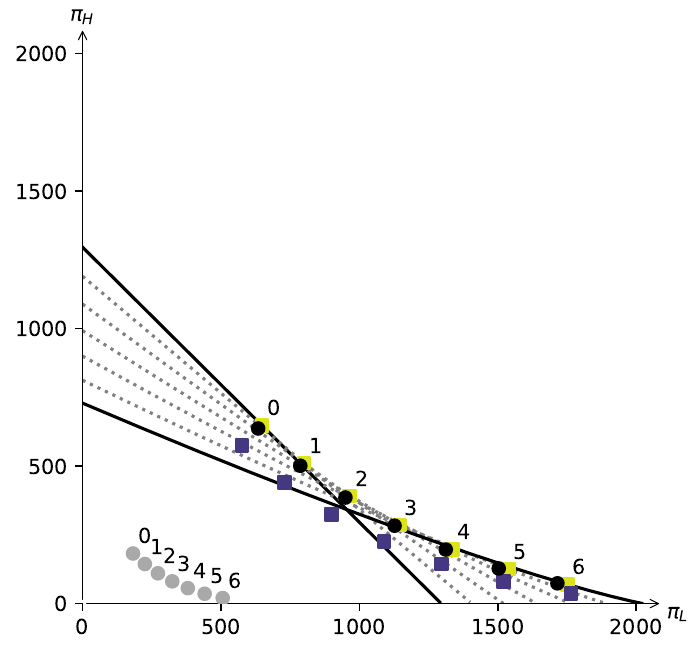}
  \caption{k=1}
  \label{fig:pareto:k1}
\end{subfigure}%
\begin{subfigure}{.5 \textwidth}
  \centering
  \includegraphics[width=0.95\linewidth]{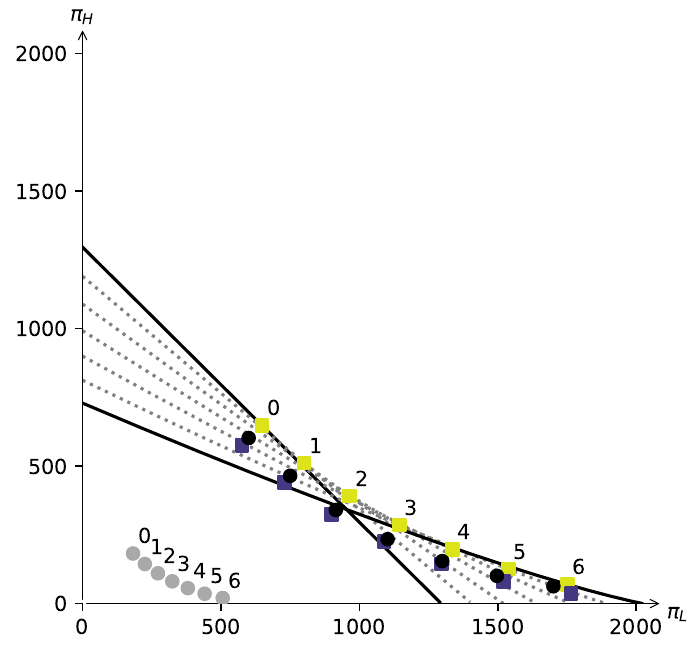}
  \caption{k=0}
   \label{fig:pareto:k0}
\end{subfigure}
\caption{Pareto frontier and firm profits for our simulation results (black circles), equal relative gains (yellow squares), Nash equilibrium (purple squares), and min-max-disagreement (grey circles).}
\label{fig:pareto}
\end{center}
\end{figure}

In Figure \ref{fig:pareto}, we shed light on how algorithms solve the allocation bargaining problem. Panel (a) shows the outcomes for our main specification, where algorithms have a memory length of $k=1$. The x-axis depicts the profits of the more efficient low-cost firm ($\Pi_L$) while the y-axis represents the profits of the inefficient high-cost firm ($\Pi_H$). 

The solid and dotted lines show the Pareto frontiers of profits for different degrees of asymmetry. The steepest solid line corresponds to the Pareto frontier of the symmetric case, while the solid line with the flattest slope corresponds to the most asymmetric case (\textit{asym6}). The dotted lines depict the Pareto frontiers of \textit{asym1} to \textit{asym5}. Additionally, the figure depicts the profits obtained in our simulations (black dots), under equal relative gains (yellow squares), Nash equilibrium (purple squares), and min-max-disagreement (grey circle) for different degrees of asymmetry ($0 = $ symmetry, $6 = $ most asymmetric). 

For all degrees of asymmetry, our results lie close to the Pareto frontier, with profits significantly exceeding those of the Nash equilibrium for both firms. Profits achieved by the algorithms are more than double those under min-max-disagreement. What is particularly striking is that the profits are close to equal relative gains for all degrees of asymmetry, confirming our results from Section \ref{sec:explanatory-power}. This suggests that the idea underlying equal relative gains are a good description of how algorithms reconcile the allocation problem: The spoils of collusion are divided such that both firms gain from collusion equally much \textit{in relative terms}, and, importantly, such that the total profits are close to the Pareto frontier.

\paragraph{Algorithmic sophistication.} A common concern in the analysis of self-learning algorithms regards the question of algorithmic sophistication, especially when algorithms obtain supra-competitive outcomes as in our setting \citep[see, for instance,][]{calvano2023algorithmic, epivent2024algorithmic}. Do the algorithms obtain high profits either because they are sufficiently sophisticated to develop strategies that support these outcomes, or are these outcomes rather a consequence precisely of the \textit{lack of sophistication}, leaving gains of exploiting an innocent cooperator on the table? 

To address these concerns, we provide two pieces of evidence. First, we investigate the outcomes that the algorithms would obtain if we do not endow them with any memory ($k=0$). This mechanically prevents the emergence of any reward-and-punishment schemes typically encountered in collusive strategies. Not allowing the firms to condition their outcomes on previous periods' actions effectively renders the game static, so that the Nash outcomes become the unique (subgame-perfect) equilibrium of the game. 

The results of these simulations are shown in panel (b) of Figure \ref{fig:pareto}. For all degrees of asymmetry ($0 = $ symmetry, $6 = $ most asymmetric), our simulation results (black circles) are very close to the Nash equilibrium outcomes (grey circles) and well below the Pareto frontier. This is in sharp contrast to the results we obtained when we endowed the agents with memory ($k=1$ in panel (a)).

To further investigate algorithmic sophistication, we provide a second piece of evidence by examining the incentives to deviate from the strategy prescribed by the Q-matrix upon convergence. Following \cite{calvano2020artificial}, for each simulation, agent $i$, and parameterization of the game, we compute the static best response to the strategy of the other player. Note that this computation is beyond the knowledge available to the algorithms and hence accessible only to us, setting the bar for algorithmic sophistication very high. It allows the deviating agent to gain the maximum one-period deviation profit.\footnote{
As an alternative deviation analysis, we let the agent choose a strictly higher quantity with the highest Q-value as their deviation action. The results of this robustness check are provided in Figure~\ref{fig:incentives-to-deviate-all} in the Appendix and are overall similar.}

To evaluate the profitability, for both approaches, we simulate that agent $i$ deviates to the respective action once. Subsequently, we simulate outcomes for 40 periods using the Q-matrices of both agents upon convergence.\footnote{As shown by \cite{calvano2020artificial}, following a deviation, after a certain number of periods the algorithms usually return to playing the actions upon convergence. Thus, by choosing 40 simulation periods, we ensure that the deviation circle is completed.} We then compute the fraction of simulations in which such a one-shot deviation was profitable. We consider a deviation profitable when the sum of discounted profits is higher under deviation compared to no deviation. For discounting, we consider the deviation period as $t=0$ and $\delta=0.95$ as in the initial Q-learning simulation. 

\begin{figure}[H]
\begin{center}
  \centering
  \includegraphics[scale=0.6]{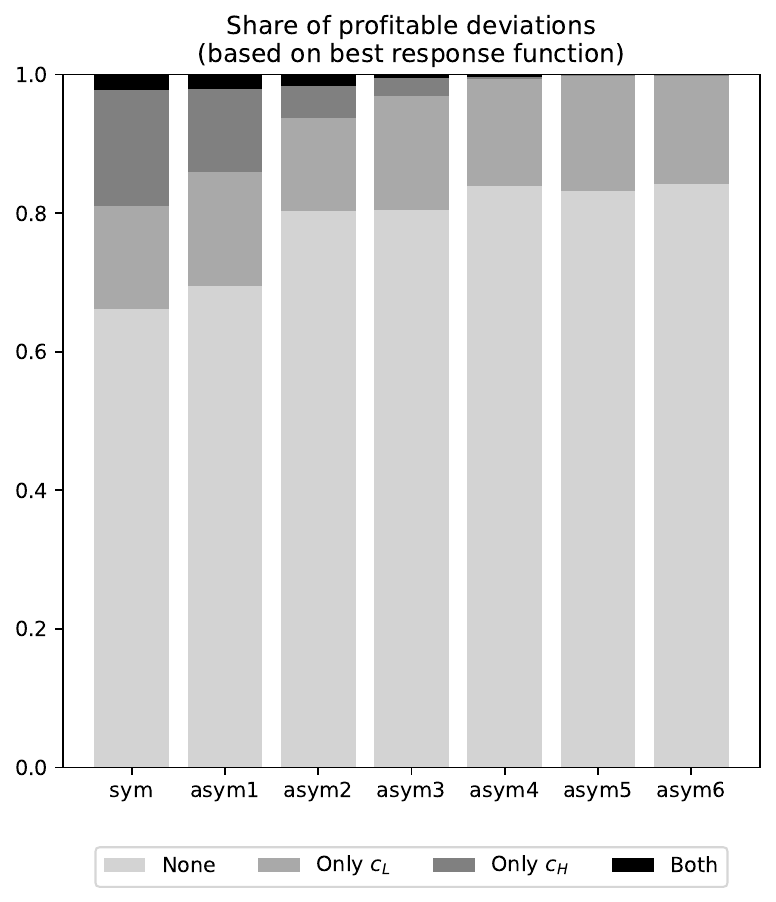}
\end{center}
\caption{Incentives to deviate based on one-period best response (moderate exploration)}
\label{fig:deviation_best_response_CV_main}
\end{figure}

The results are shown in Figure \ref{fig:deviation_best_response_CV_main}. For all degrees of asymmetry, in more than 60\% of cases, none of the agents has any incentive to deviate. Only rarely do both agents want to deviate, but in this specification, the cost-efficient $c_L$ firm might want to deviate in around 15\% of the simulations.

Taken together, we interpret the two separate considerations of memory and deviation as very suggestive of the fact that collusive outcomes observed in our setting are indeed the result of sophisticated algorithms that learned to operate effectively in their environment.

\subsection{Robustness checks}
\label{sec:robustness-checks}

\paragraph{Alternative cost asymmetry.}
The cost parameterizations of our main specification generally only depicts one of infinitely many possible cost asymmetries. Although chosen thoughtfully by keeping constant total Nash quantities, we provide an alternative cost parameterization in Appendix~\ref{sec:robustness-cost-asymmetry}. Under the alternative cost parameterization, the one-firm monopoly output is kept constant (that is, $c_L=19$) while we increase the costs of the high-cost firm $c_H$. For this exercise, we focus on moderate exploration ($\nu=21$). Qualitatively, the results are similar to our main specification. Firms reduce total quantities well below the Nash prediction and the profit split lies on the Pareto frontier for all cases. The profit split aligns closely with the equal relative gains prediction using min-max deviation profits but exhibits some deviations for high degrees of asymmetry. As in our main specification, Nash and equal relative gains (particularly with Nash deviation profits) effectively capture the relative changes compared to the symmetric case.

\paragraph{Different $\alpha, \beta$ parameters.}
In our main specification, we consider moderate exploration such that the expected times of random visits per cell in the Q-matrix is $\nu=21$.
We assumed an $\alpha=0.15$. As a robustness test, we ran simulations with high exploration ($\nu=100$), leading to qualitatively similar results, as shown in Figures~\ref{fig:res:1_all}, \ref{fig:res:comp:full}, \ref{fig:main_k0:res:1}, and \ref{fig:incentives-to-deviate-all} in the Appendix. For simulation results with high exploration, we only observe a slightly stronger increase of total quantities for higher degrees of asymmetry and an even lower incentive to deviate compared to moderate exploration.
As a robustness check, we further investigate the outcomes when we set the learning parameter $\alpha = 0.2$ instead for moderate and high exploration. Our results remain robust to this adjustment as shown in Figure~\ref{fig:res:alpha:1} and \ref{fig:pareto:alpha} in the Appendix.

\paragraph{Algorithmic sophistication.} 
Our deviation analysis in Figure~\ref{fig:deviation_best_response_CV_main} shows an increase in the share of simulation runs where the deviation was not profitable. Given that we also observe an increase in total quantities for higher degrees of asymmetry, one might suspect the share of profitable deviations to be the main driver of this effect. To address this, we evaluated total quantities separately for two subsamples: (i) incentive-compatible simulations and (ii) simulations where at least one firm could profitably deviate. The robustness check reveals the same trend for both subsamples, with average total quantities in incentive-compatible simulations approximately 0.5 units higher than in simulations where at least one firm could profitably deviate, see Figure~\ref{fig:quantities:subsamples}.

\section{Conclusion}
\label{sec:Conclusion}
Sustaining collusion in asymmetric industries poses a significant challenge, as firms must tacitly agree on both total output and profit allocation. While an equal profit split is an intuitive focal point in symmetric markets, asymmetries introduce complexities, raising questions about whether collusion can be sustained and how firms allocate profits.

This paper complements the existing theoretical, empirical, and laboratory literature on collusion in asymmetric markets by conducting several simulation experiments with self-learning algorithms. In particular, we let the self-learning algorithms compete in a Cournot duopoly and gradually increase the degree of cost asymmetry while keeping the total (static) Nash quantity constant. The approach allows us to benchmark the different degrees of asymmetry against a robust counterfactual to illustrate what a collusive outcome might look like. 

Our results reveal that algorithms consistently reduce total output below the static Nash equilibrium across all degrees of asymmetry, indicating tacit collusion. For a higher degree of asymmetry, the results imply an increase in total quantities, and thus consumer surplus. While total profits in our simulation exceed total Nash profits for lower degrees of asymmetry, the effect diminishes for higher degrees of asymmetry. Nevertheless, this does not necessarily imply a lower degree of (tacit) collusion, as firm profits for all degrees of asymmetry lie on the Pareto frontier. Thus, the firm algorithms robustly established cooperation.

We evaluate the simulation results against several different theoretical predictions. The concept of equal relative gains is the bargaining idea that best captures the effect of asymmetry and overall results. Firms allocate cooperative profits in \textit{relative} terms to a disagreement point.

Our findings have important implications for competition policy. Typically, symmetry among firms is viewed as potentially facilitating collusion and, \textit{ceteris paribus}, demanding the competition authority's closest attention. However, according to our results, the opposite may hold in industries where decisions are increasingly delegated to autonomous algorithms: precisely \textit{asymmetric} firms might be more prone to dampening competition and yielding inefficient market outcomes.

\newpage

\singlespacing

\bibliographystyle{aer}

\bibliography{refs.bib}

\newpage

\pagebreak

\appendix
\renewcommand{\thesubsection}{\Alph{subsection}}
\counterwithin{figure}{section}
\counterwithin{table}{section}

\section*{Appendix}

\section{Additional Tables}
\label{sec:additional-tables}

\begin{table}[htb!]
\centering
\caption{Average squared normalized distances (multiplied by 1,000) from simulation results (moderate exploration)}
\label{tab:average_distances_normalized}
\begin{tabular}{lrrrr}
\hline
 & Q & CS & $\Pi$ & W \\
\hline
Nash & 0.34 & 1.41 & 5.81 & 0.03 \\
Monopoly & 18.78 & 92.02 & 18.63 & 39.95 \\
Alternating Monopoly & 0.34 & 1.41 & 30.38 & 13.86 \\
Kalai-Smorodinsky & 3.44 & 13.21 & 29.80 & 20.60 \\
Equal Split & 12.30 & 43.38 & 97.45 & 71.79 \\
Equal Relative Gains & 0.86 & 3.71 & 0.01 & 0.81 \\
Kalai-Smorodinsky (Nash) & 0.98 & 4.31 & 0.04 & 1.05 \\
Equal Relative Gains (Nash) & 2.71 & 12.28 & 1.03 & 4.59 \\
\hline
\end{tabular}
\end{table}

\newpage

\section{Additional Figures}
\label{sec:additional-figures}

\begin{figure}[hbt!]
\begin{center}
\begin{subfigure}{.5 \textwidth}
  \centering
  \includegraphics[width=0.95\linewidth]{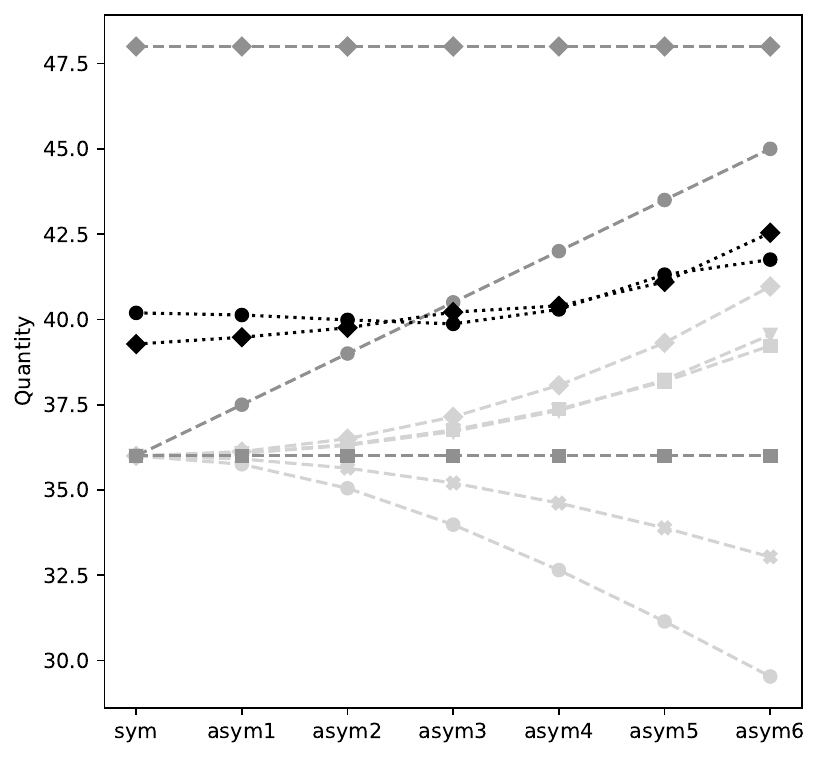}
  \caption{Total Quantities}
  \label{fig:level:q_all}
\end{subfigure}%
\begin{subfigure}{.5 \textwidth}
  \centering
  \includegraphics[width=0.95\linewidth]{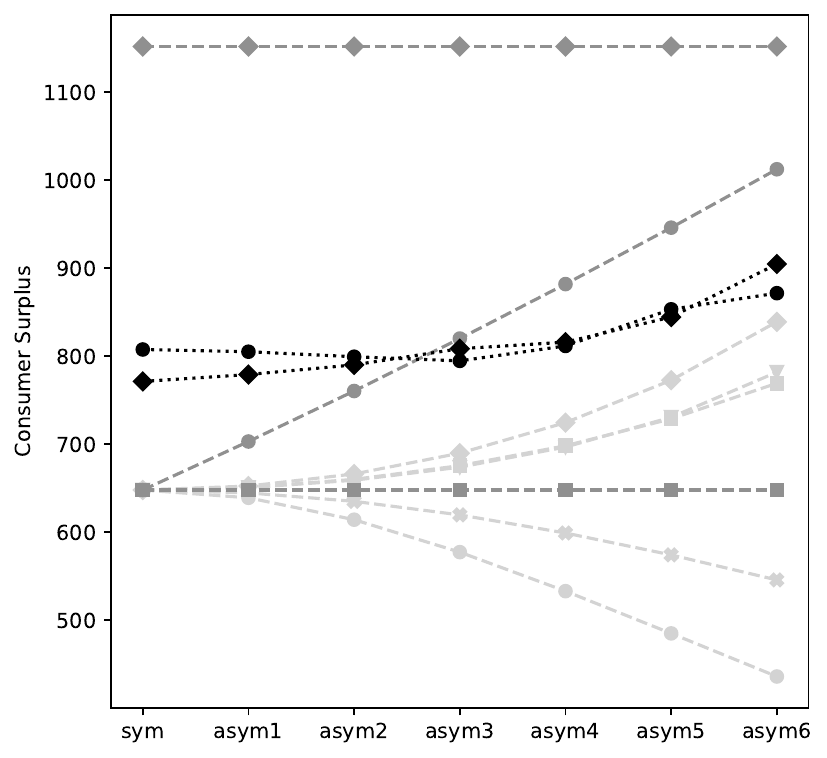}
  \caption{Consumer Surplus}
   \label{fig:level:cs_all}
\end{subfigure}
\begin{subfigure}{.5 \textwidth}
  \centering
  \includegraphics[width=0.95\linewidth]{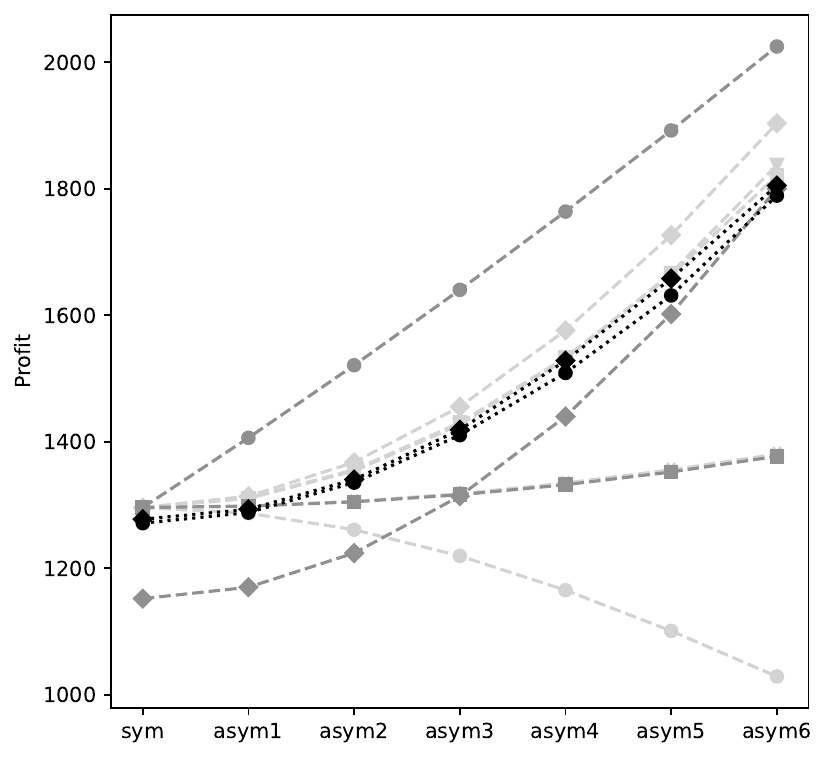}
  \caption{Total Profits}
  \label{fig:level:pi_all}
\end{subfigure}%
\begin{subfigure}{.5 \textwidth}
  \centering
  \includegraphics[width=0.95\linewidth]{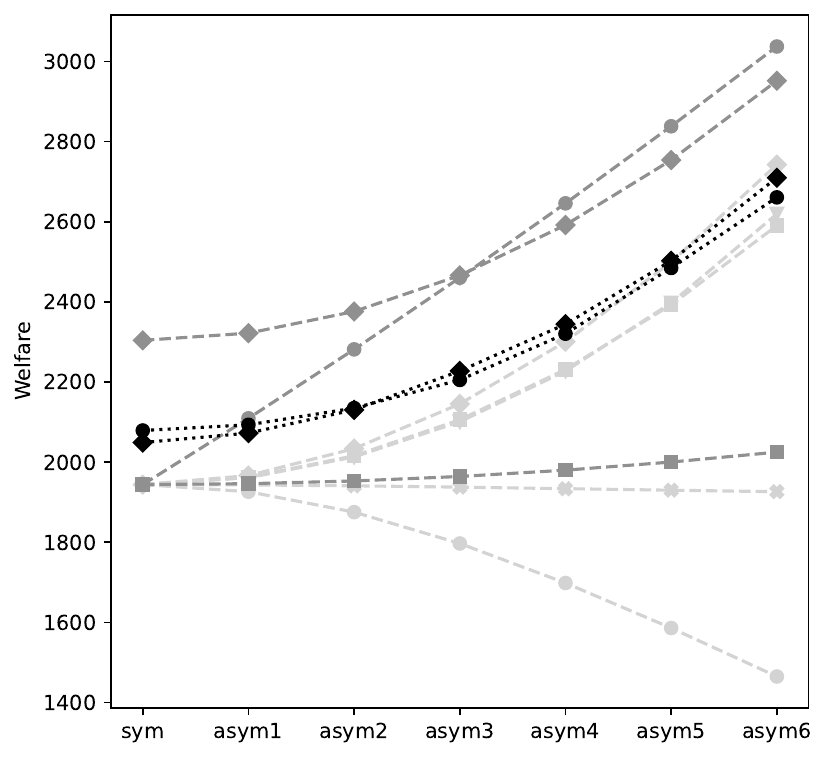}
  \caption{Total Welfare}
\end{subfigure}
\begin{subfigure}{.95 \textwidth}
  \centering
  \includegraphics[width=0.85\linewidth]{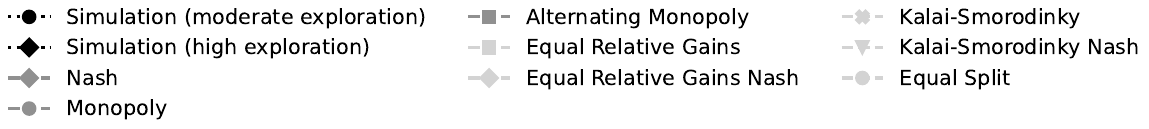}
\end{subfigure}
\caption{Simulation results for different degrees of asymmetry (main results including high exploration; all theoretical predictions).}
\label{fig:res:1_all}
\end{center}
\end{figure}

\begin{figure}[hbt!]
\begin{center}
\begin{subfigure}{.5 \textwidth}
  \centering
  \includegraphics[width=0.95\linewidth]{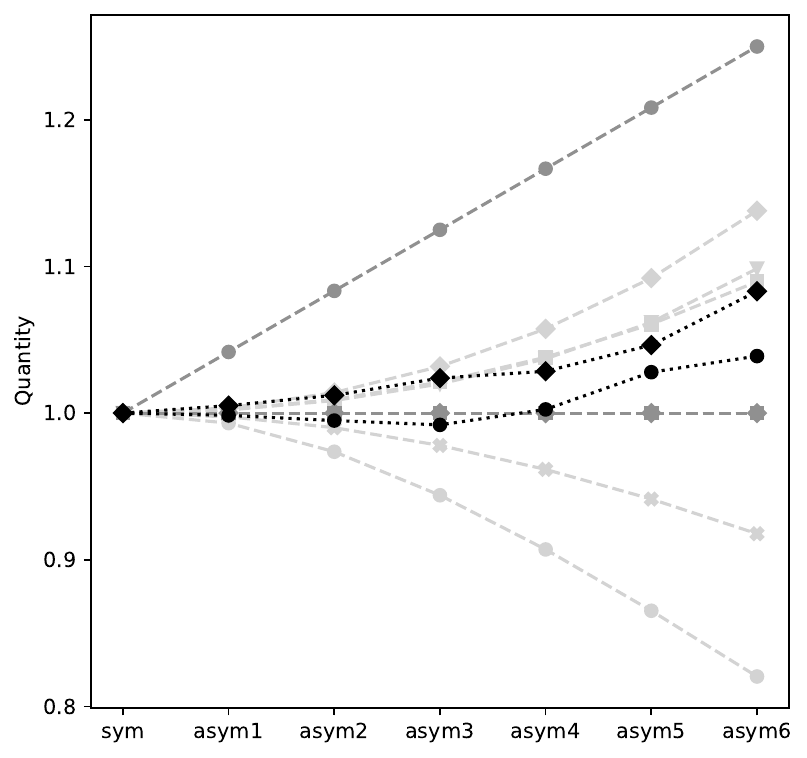}
  \caption{Total Quantities}
  \label{fig:comp:full:q}
\end{subfigure}%
\begin{subfigure}{.5 \textwidth}
  \centering
  \includegraphics[width=0.95\linewidth]{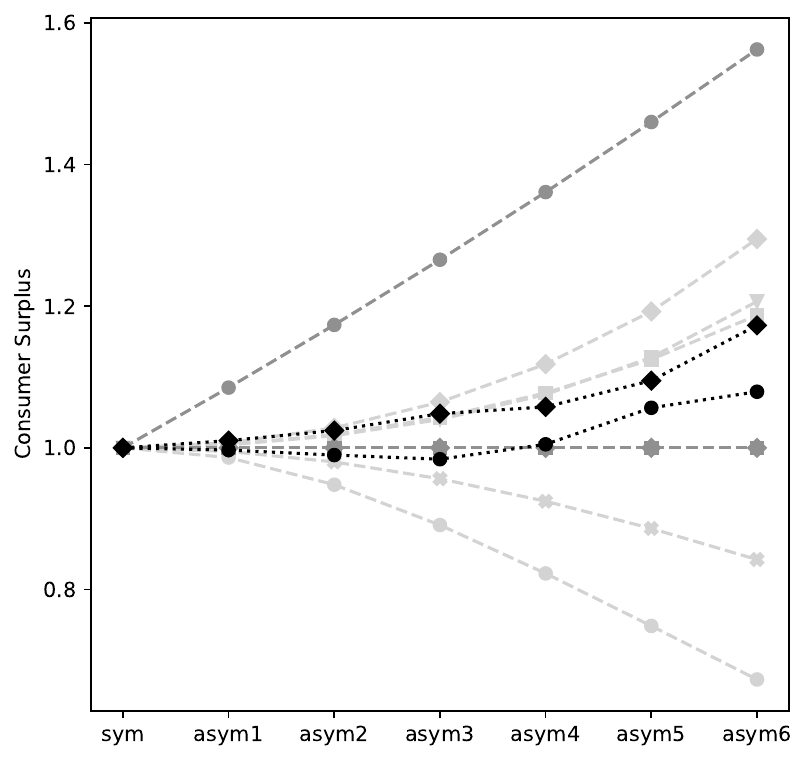}
  \caption{Consumer Surplus}
   \label{fig:comp:full:cs}
\end{subfigure}
\begin{subfigure}{.5 \textwidth}
  \centering
  \includegraphics[width=0.95\linewidth]{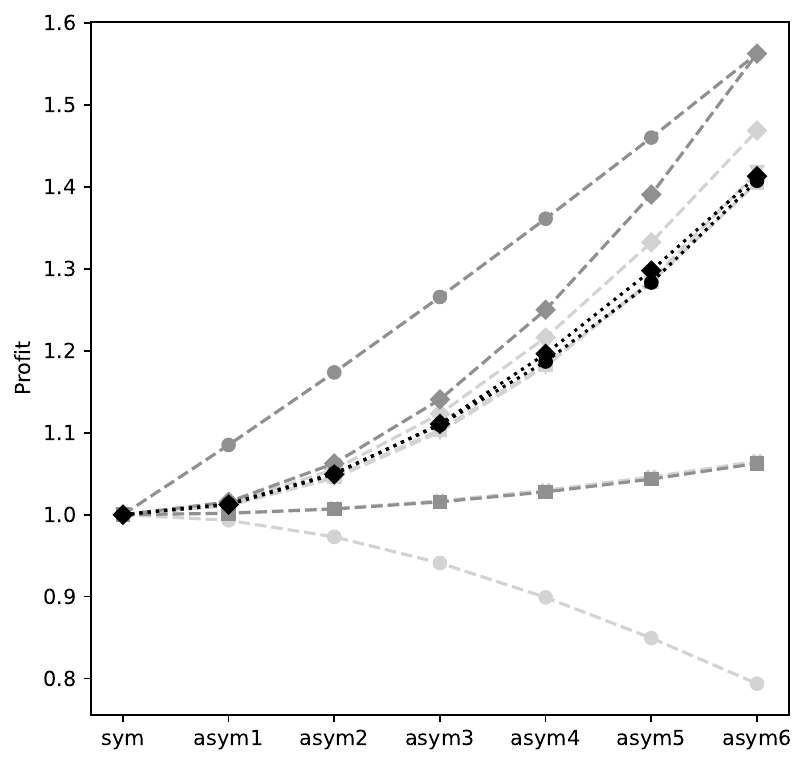}
  \caption{Total Profits}
  \label{fig:comp:full:pi}
\end{subfigure}%
\begin{subfigure}{.5 \textwidth}
  \centering
  \includegraphics[width=0.95\linewidth]{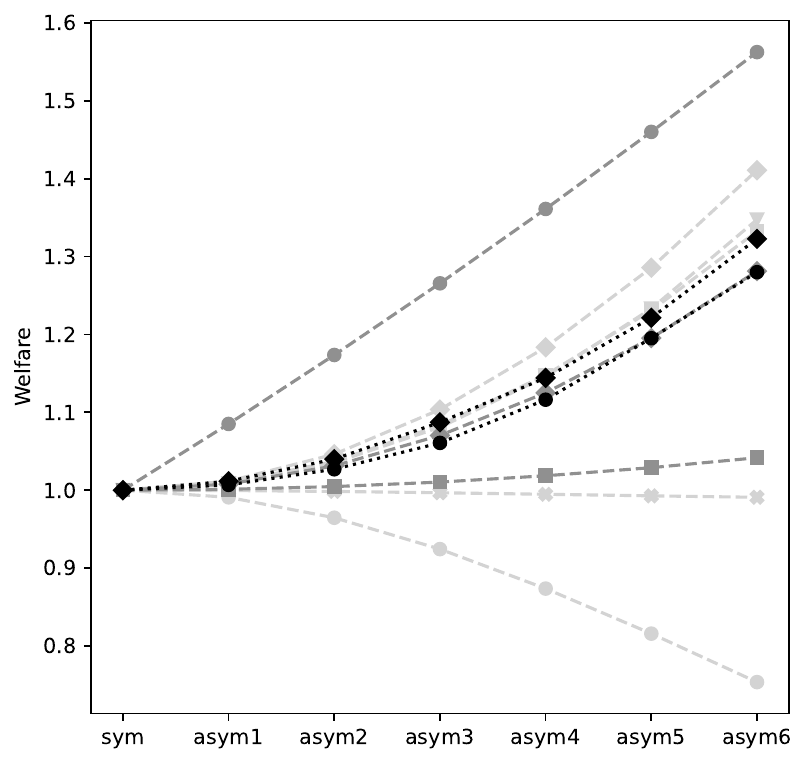}
  \caption{Total Welfare}
   \label{fig:comp:full:w}
\end{subfigure}
\begin{subfigure}{.95 \textwidth}
  \centering
  \includegraphics[width=0.85\linewidth]{img/legend.pdf}
\end{subfigure}
\caption{Simulation results (comparative statics; symmetric case = 1) for different degrees of asymmetry. }
\label{fig:res:comp:full}
\end{center}
\end{figure}

\begin{figure}[hbt!]
\begin{center}
\begin{subfigure}{.5 \textwidth}
  \centering
  \includegraphics[width=0.95\linewidth]{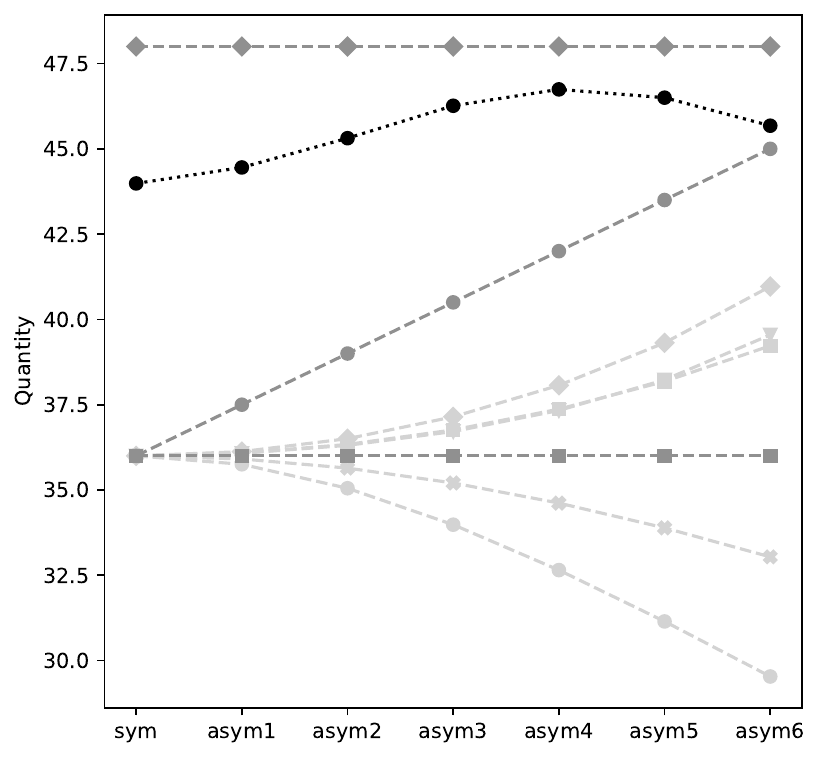}
  \caption{Total Quantities}
  \label{fig:main_k0:level:q}
\end{subfigure}%
\begin{subfigure}{.5 \textwidth}
  \centering
  \includegraphics[width=0.95\linewidth]{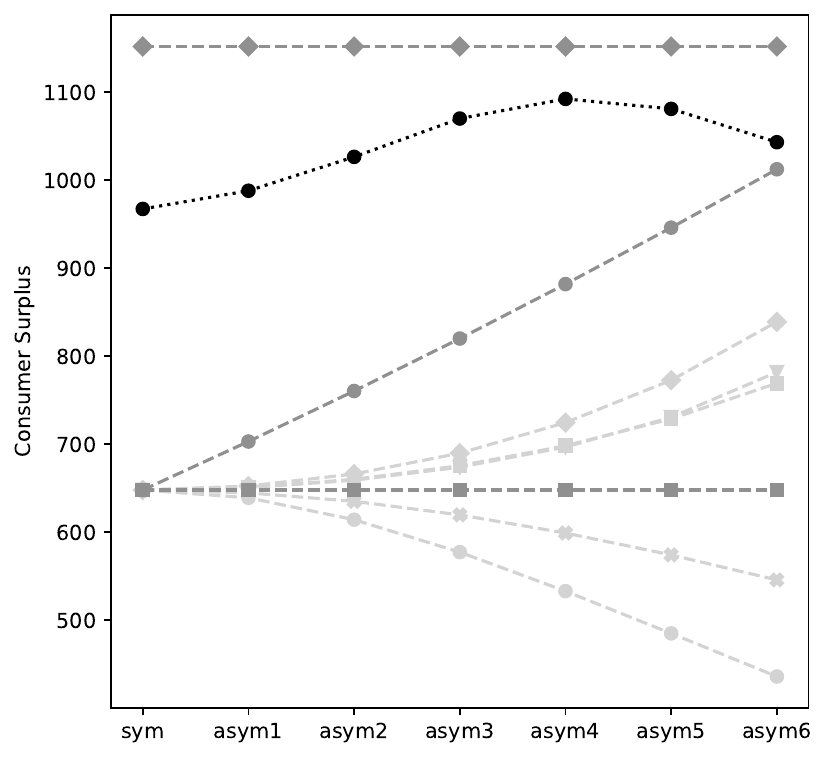}
  \caption{Consumer Surplus}
   \label{fig:main_k0:level:cs}
\end{subfigure}
\begin{subfigure}{.5 \textwidth}
  \centering
  \includegraphics[width=0.95\linewidth]{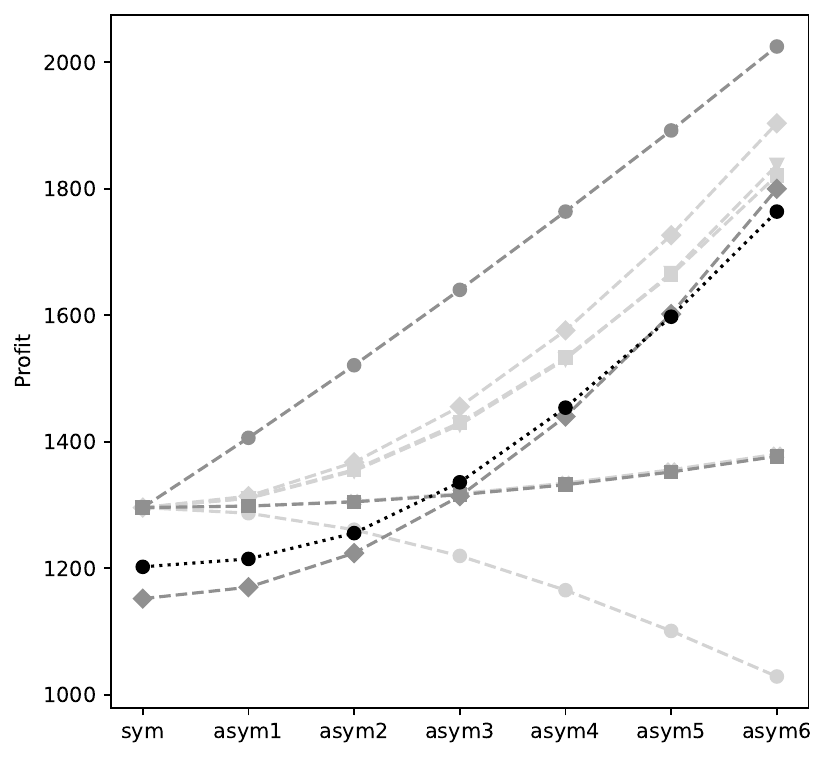}
  \caption{Total Profits}
  \label{fig:main_k0:level:pi}
\end{subfigure}%
\begin{subfigure}{.5 \textwidth}
  \centering
  \includegraphics[width=0.95\linewidth]{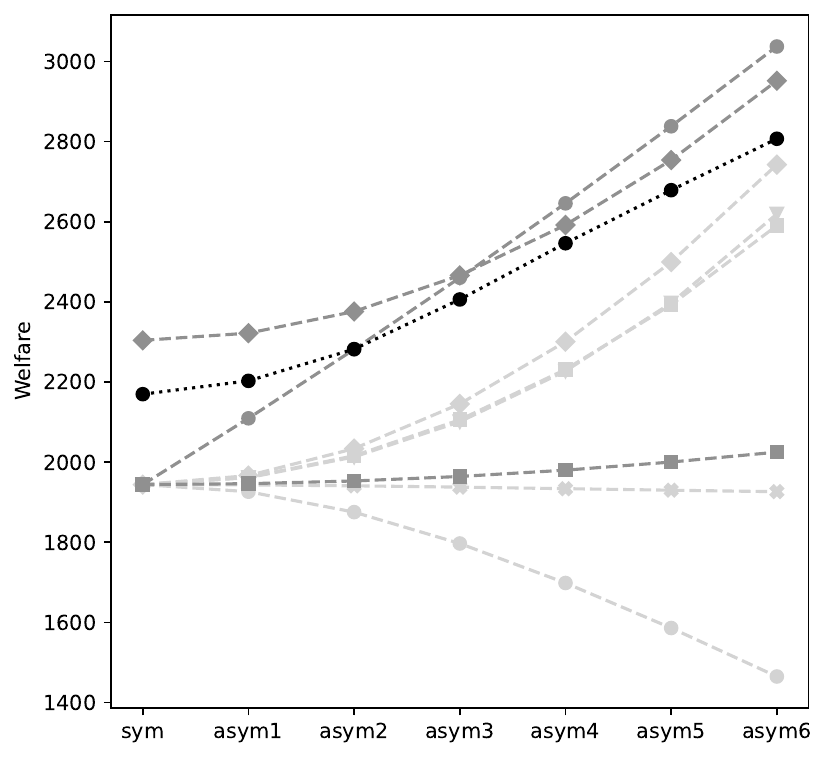}
  \caption{Total Welfare}
\end{subfigure}
\begin{subfigure}{.95 \textwidth}
  \centering
  \includegraphics[width=0.85\linewidth]{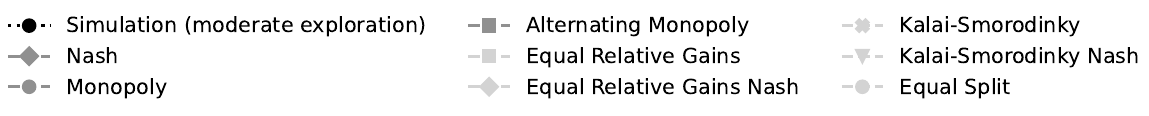}
\end{subfigure}
\caption{Simulation results (levels) for different degrees of asymmetry without memory.}
\label{fig:main_k0:res:1}
\end{center}
\end{figure}

\begin{figure}[hbt!]
\begin{center}
\begin{subfigure}{.5 \textwidth}
  \centering
  \includegraphics[width=0.95\linewidth]{img/deviation_best_response_CV.pdf}
  \caption{Best response (moderate exploration)}
  \label{fig:deviation_best_response_CV}
\end{subfigure}%
\begin{subfigure}{.5 \textwidth}
  \centering
  \includegraphics[width=0.95\linewidth]{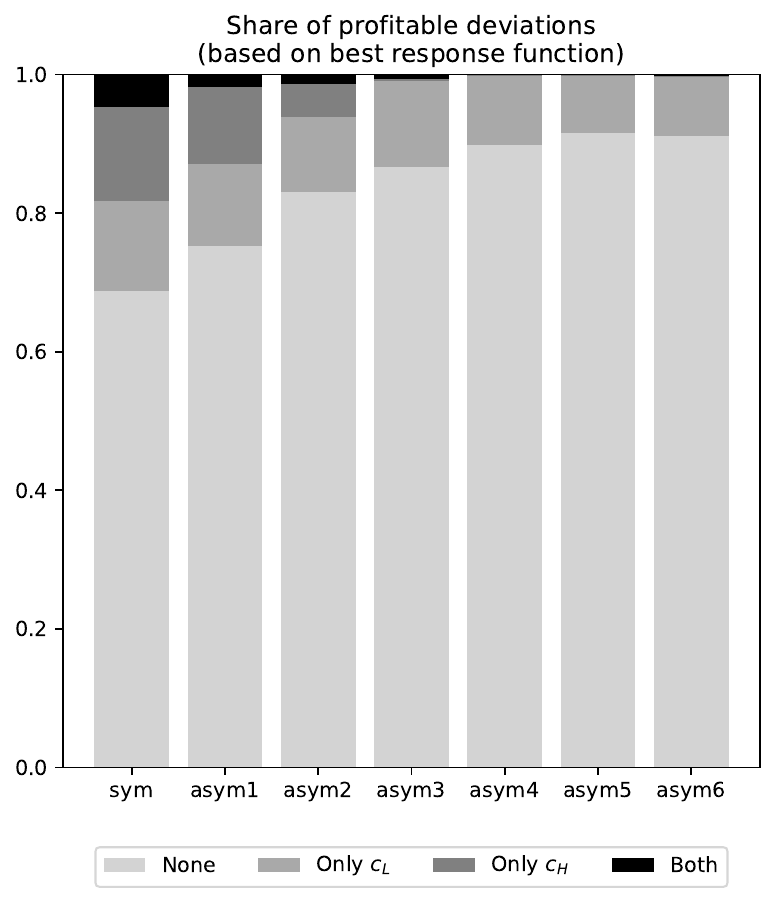}
  \caption{Best response (high exploration)}
  \label{fig:deviation_best_response_low_beta}
\end{subfigure}
\begin{subfigure}{.5 \textwidth}
  \centering
  \includegraphics[width=0.95\linewidth]{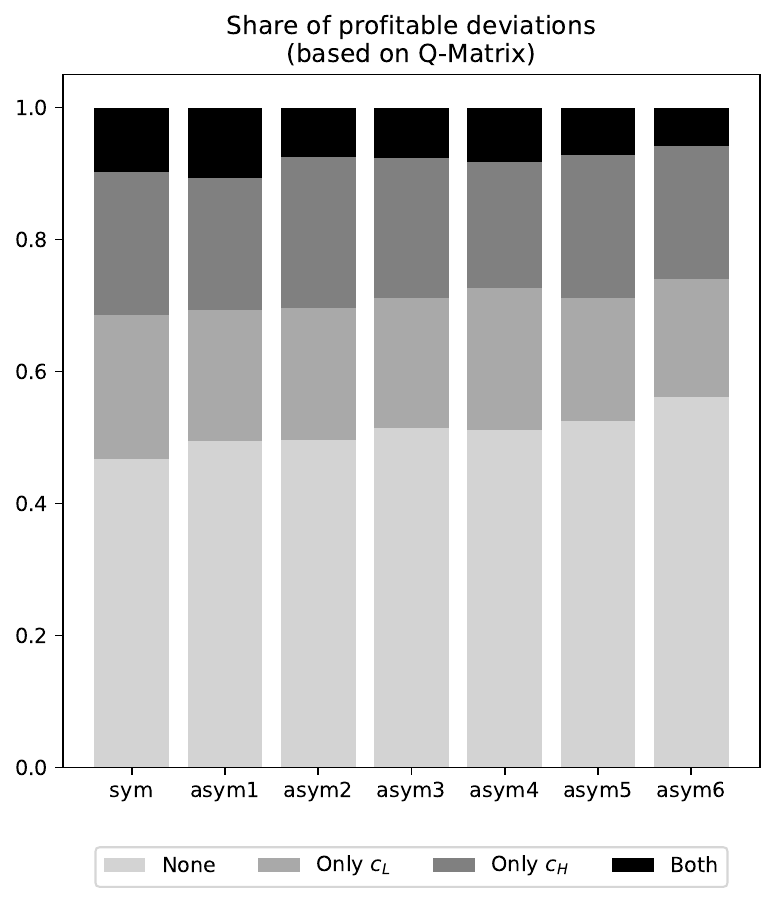}
  \caption{Q-Value (moderate exploration)}
  \label{fig:deviation_second_best_CV}
\end{subfigure}%
\begin{subfigure}{.5 \textwidth}
  \centering
  \includegraphics[width=0.95\linewidth]{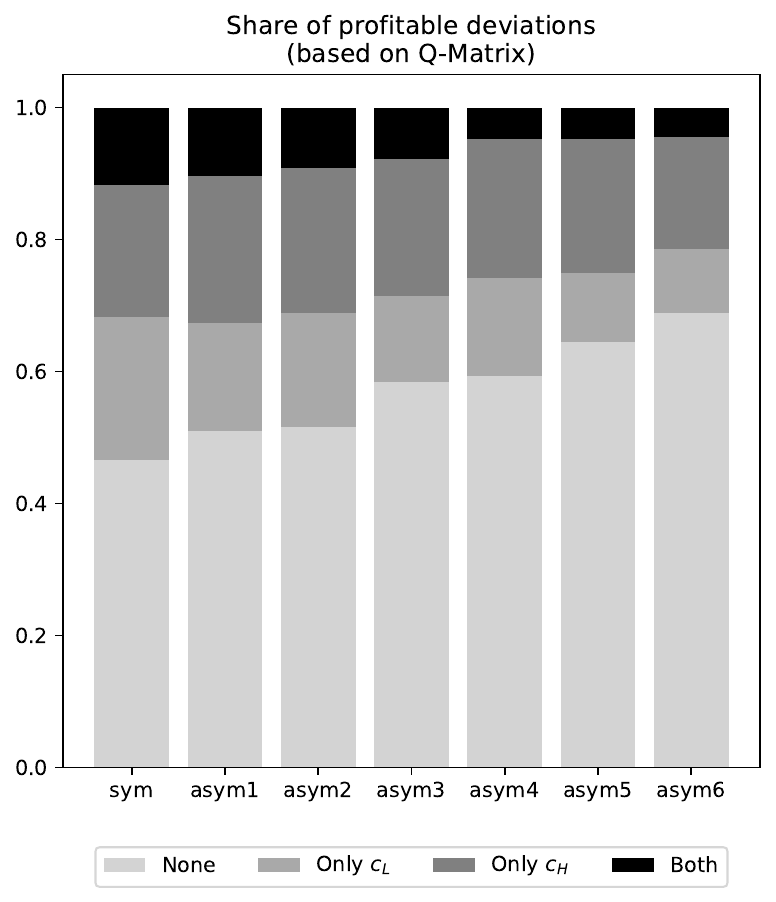}
  \caption{Q-Value (high exploration)}
  \label{fig:deviation_second_best_low_beta}
\end{subfigure}
\caption{Incentives to deviate}
\label{fig:incentives-to-deviate-all}
\end{center}
\end{figure}

\begin{figure}[hbt!]
\begin{center}
\begin{subfigure}{.5 \textwidth}
  \centering
  \includegraphics[width=0.95\linewidth]{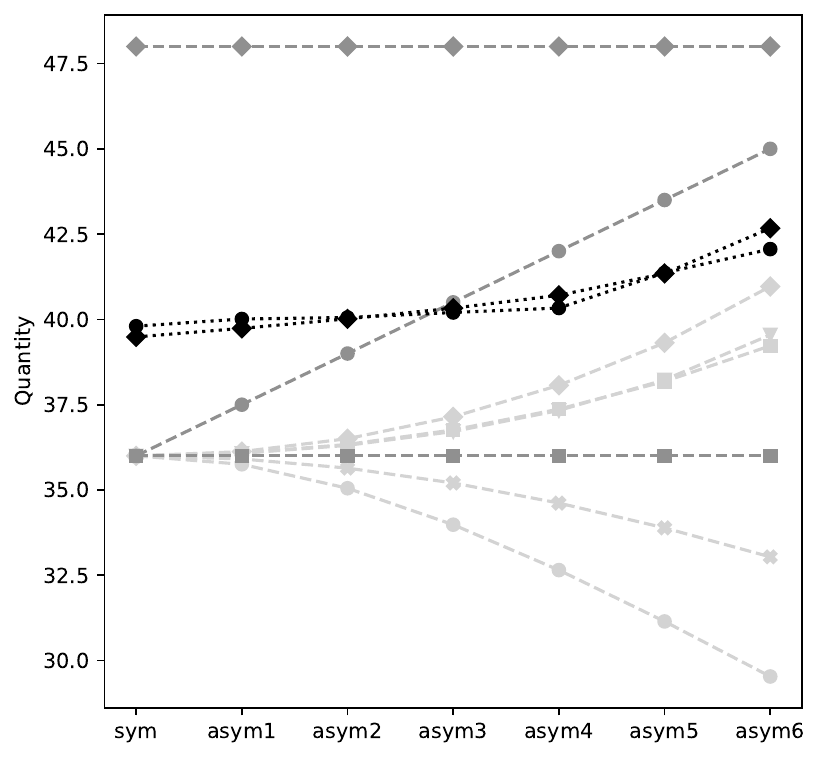}
  \caption{Total Quantities}
  \label{fig:level:alpha:q}
\end{subfigure}%
\begin{subfigure}{.5 \textwidth}
  \centering
  \includegraphics[width=0.95\linewidth]{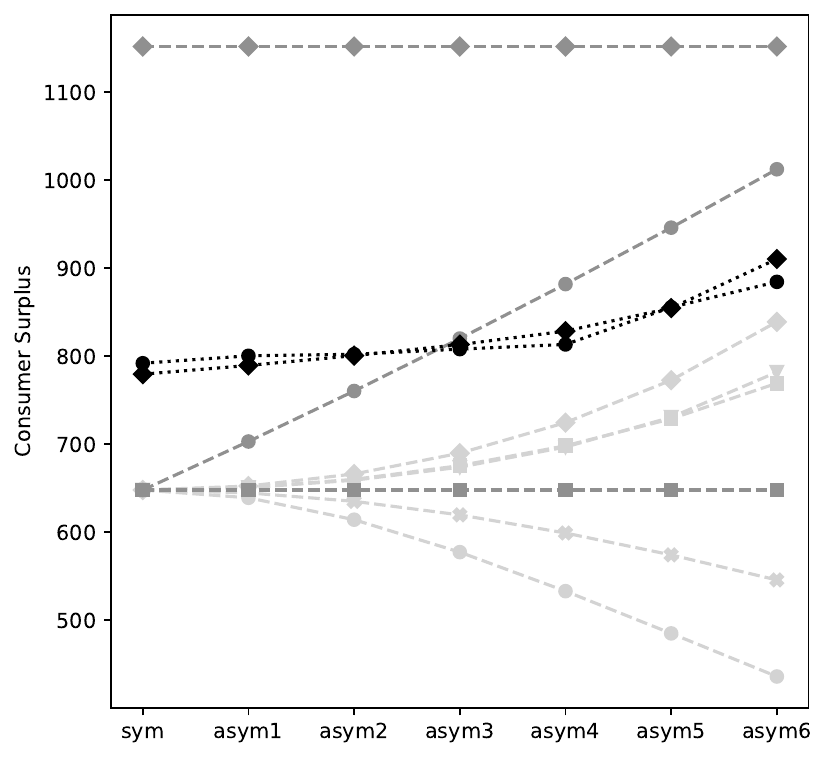}
  \caption{Consumer Surplus}
   \label{fig:level:alpha:cs}
\end{subfigure}
\begin{subfigure}{.5 \textwidth}
  \centering
  \includegraphics[width=0.95\linewidth]{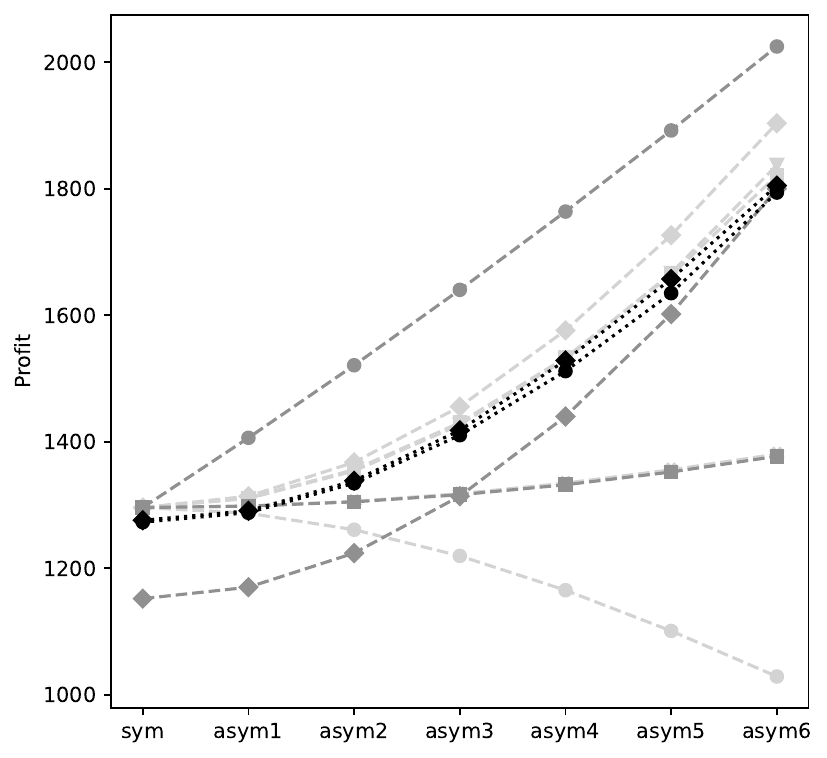}
  \caption{Total Profits}
  \label{fig:level:alpha:pi}
\end{subfigure}%
\begin{subfigure}{.5 \textwidth}
  \centering
  \includegraphics[width=0.95\linewidth]{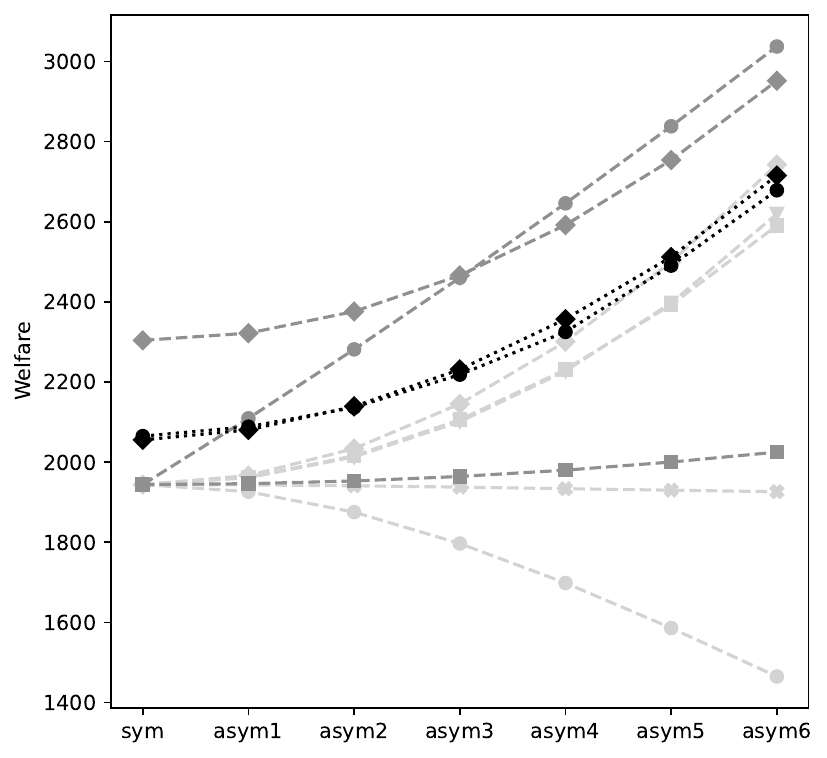}
  \caption{Total Welfare}
\end{subfigure}
\begin{subfigure}{.95 \textwidth}
  \centering
  \includegraphics[width=0.85\linewidth]{img/legend.pdf}
\end{subfigure}
\caption{Simulation results for different degrees of asymmetry ($\alpha=0.2$ including high exploration; all theoretical predictions).}
\label{fig:res:alpha:1}
\end{center}
\end{figure}

\begin{figure}[hbt!]
\begin{center}

\centering
\includegraphics[width=0.55\linewidth]{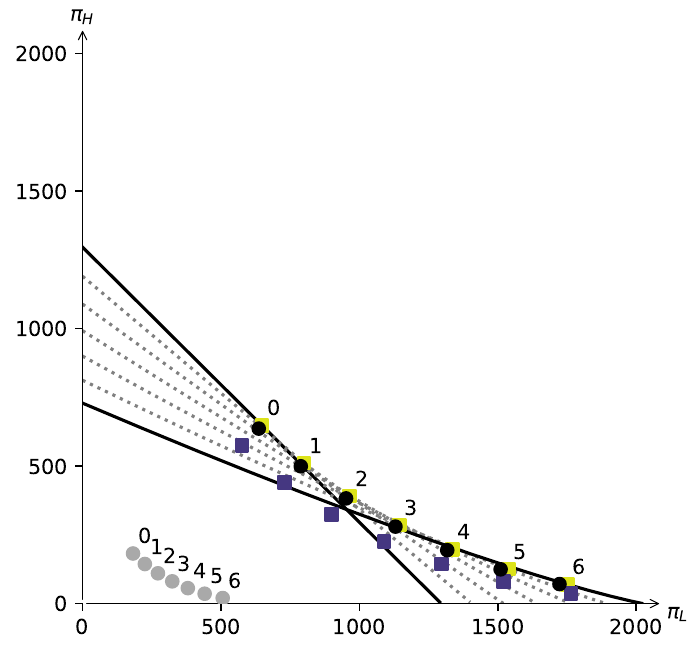}
\caption{Pareto frontier and simulation results for $\alpha=0.2$ and $\nu=21$.}
\label{fig:pareto:alpha}

\end{center}
\end{figure}

\begin{figure}[hbt!]
\begin{center}

\centering
\includegraphics[width=0.55\linewidth]{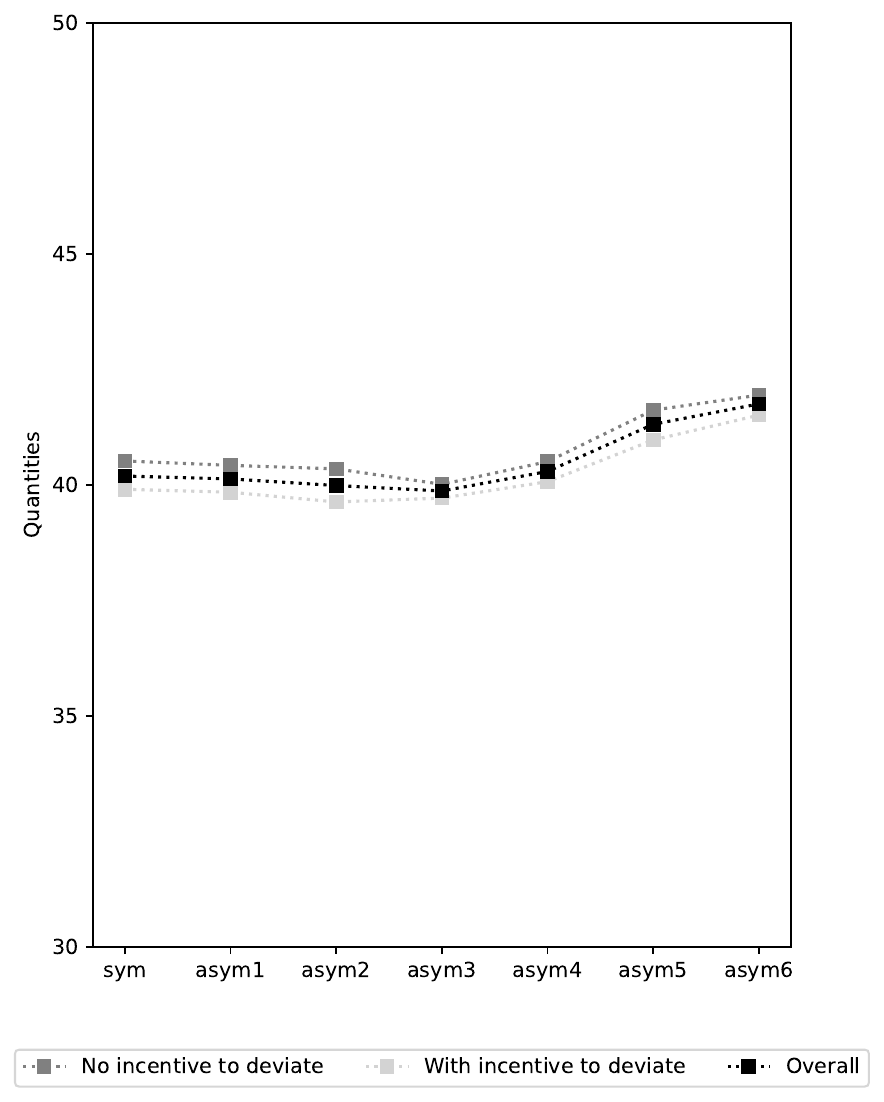}
\caption{Total quantities for main specification ($\nu=21$, $\alpha=0.15$) overall and subsamples divided based on whether deviation is profitable.}
\label{fig:quantities:subsamples}

\end{center}
\end{figure}

\FloatBarrier

\newpage

\section{Robustness Check: Alternative Cost Asymmetry }
\label{sec:robustness-cost-asymmetry}
\begin{table}[htb!]
\centering
\caption{Alternative cost parameterization}
\label{tab:costasymmetry:alt}
\begin{tabular}{lccccccc}
\hline
    & Sym & Asym1 & Asym2 & Asym3 & Asym4 & Asym5 & Asym6  \\
    \hline 
    $c_L$       & 19 & 19 & 19 & 19 & 19 & 19 & 19  \\
    $c_H$       & 19 & 22 & 25 & 28 & 31 & 34 & 37 \\
    $q_L^{NE}$  & 24 & 25 & 26 & 27 & 28 & 29  & 30 \\
    $q_H^{NE}$  & 24 & 22 & 20 & 18 & 16 & 14 & 12   \\
    $Q^{NE}$    & 48 & 47 & 46 & 45 & 44 & 43 & 42 \\
    $Q^M$       & 36 & 36 & 36 & 36 & 36 & 36 & 36 \\
\hline
\end{tabular}
\end{table}

\begin{figure}[H]
\begin{center}
\begin{subfigure}{.5 \textwidth}
  \centering
  \includegraphics[width=0.95\linewidth]{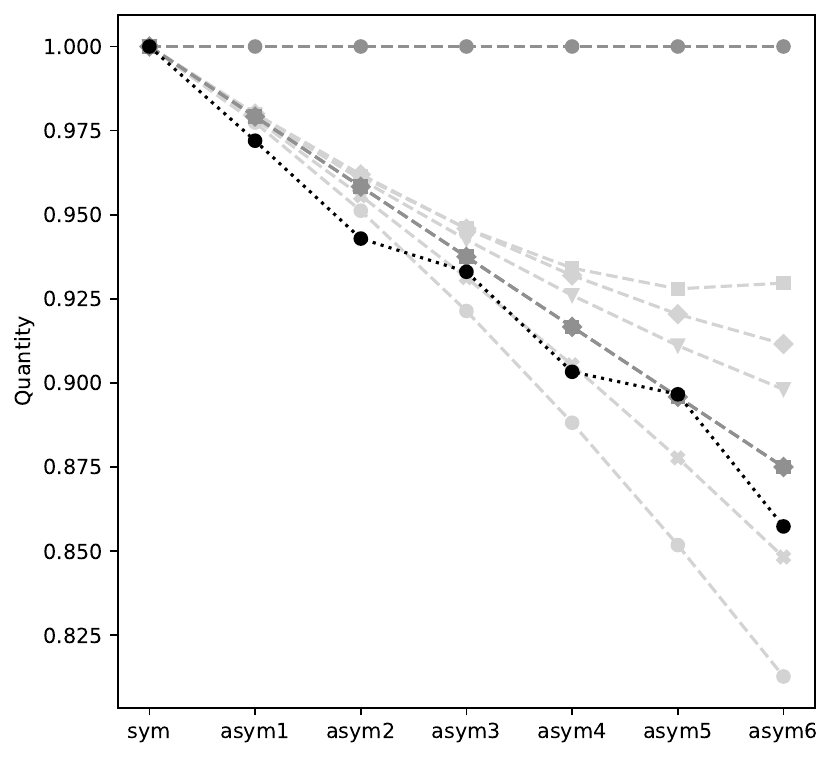}
  \caption{Total Quantities}
  \label{fig:alt_k1:comp:q}
\end{subfigure}%
\begin{subfigure}{.5 \textwidth}
  \centering
  \includegraphics[width=0.95\linewidth]{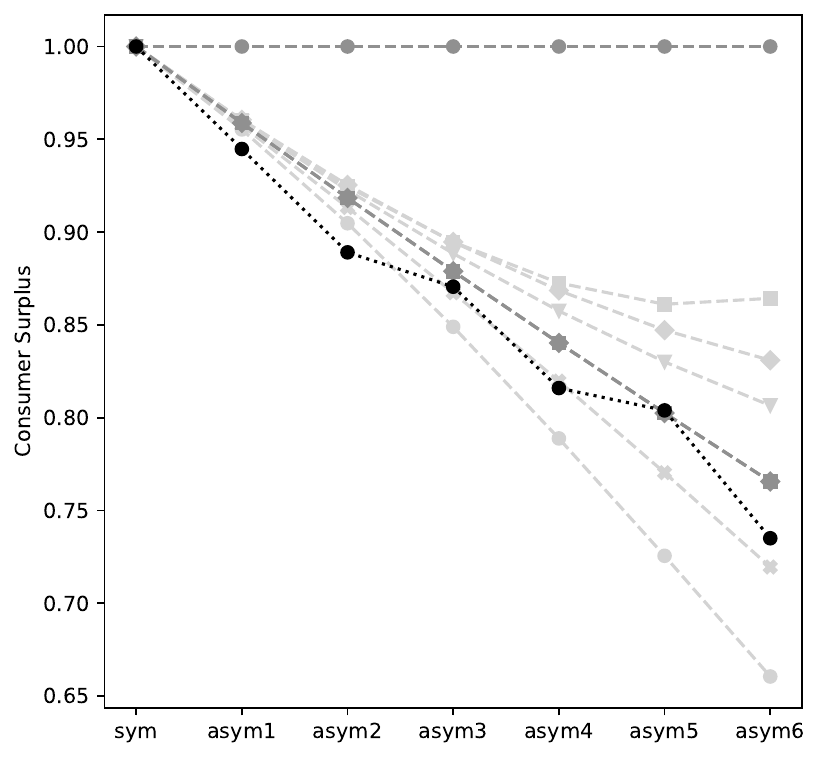}
  \caption{Consumer Surplus}
   \label{fig:alt_k1:comp:cs}
\end{subfigure}
\begin{subfigure}{.5 \textwidth}
  \centering
  \includegraphics[width=0.95\linewidth]{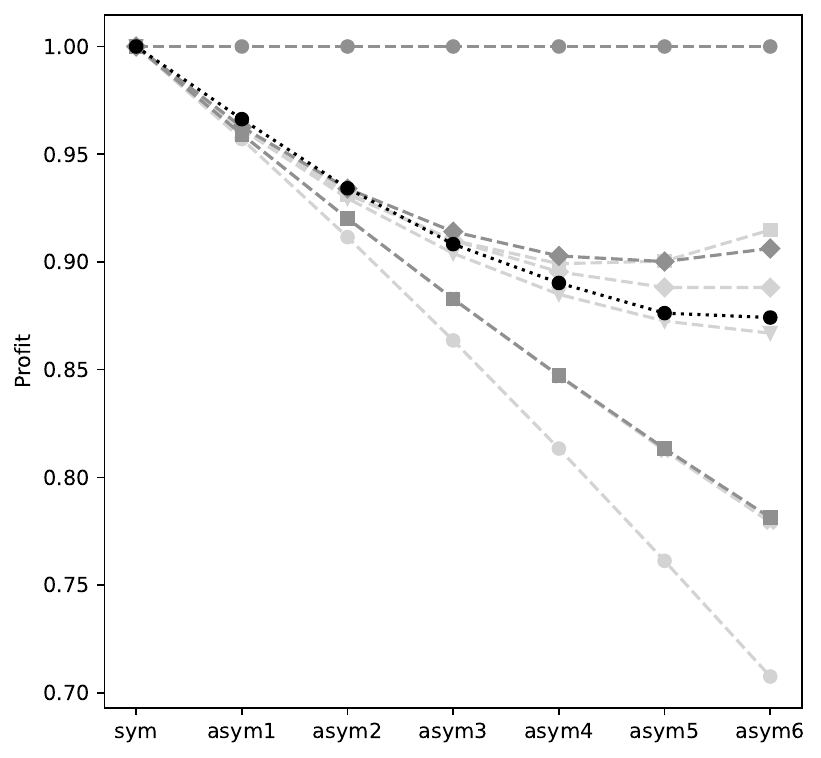}
  \caption{Total Profits}
  \label{fig:alt_k1:comp:pi}
\end{subfigure}%
\begin{subfigure}{.5 \textwidth}
  \centering
  \includegraphics[width=0.95\linewidth]{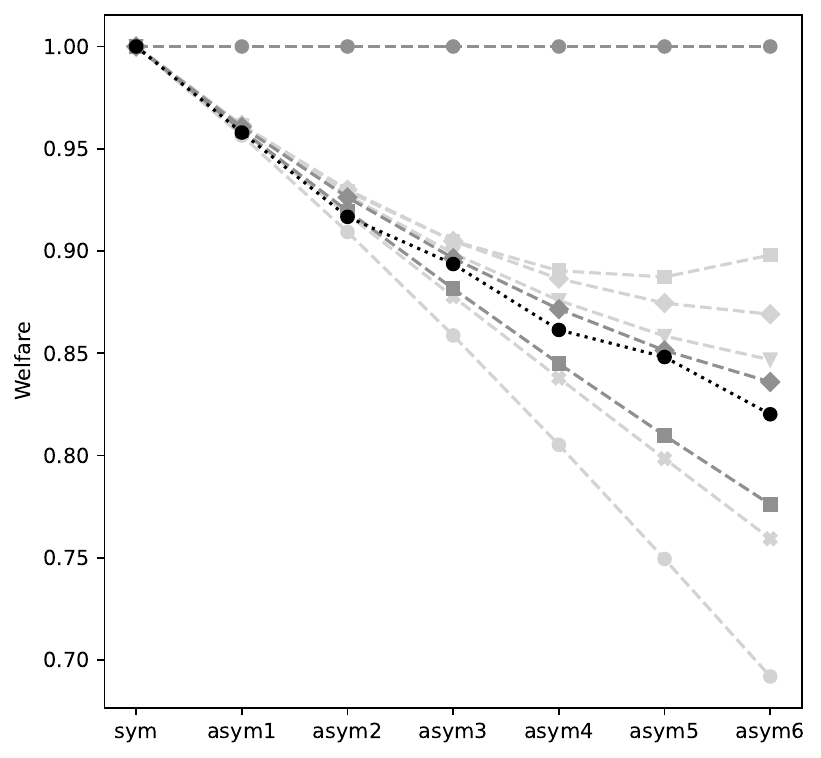}
  \caption{Total Welfare}
   \label{fig:alt_k1:comp:w}
\end{subfigure}
\begin{subfigure}{.95 \textwidth}
  \centering
  \includegraphics[width=0.85\linewidth]{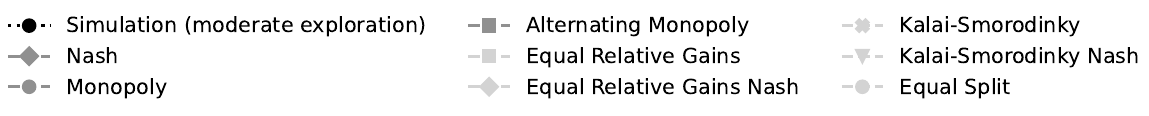}
\end{subfigure}
\caption{Simulation results (comparative statics; symmetric case = 1) for alternative different degrees of asymmetry.}
\label{fig:alt_k1:res:comp}
\end{center}
\end{figure}

\begin{figure}[H]
\begin{center}
\begin{subfigure}{.5 \textwidth}
  \centering
  \includegraphics[width=0.95\linewidth]{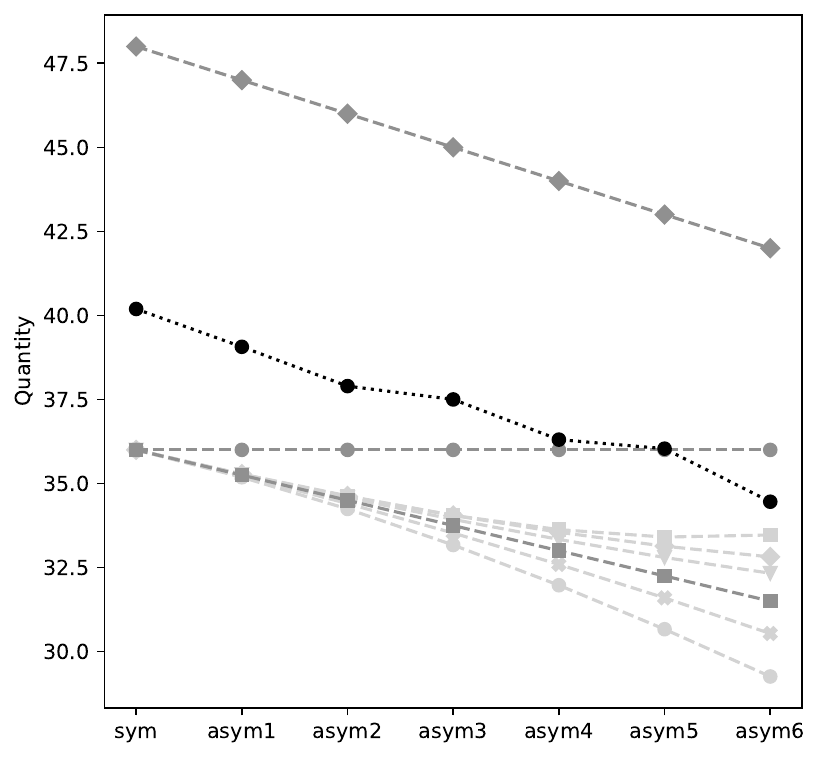}
  \caption{Total Quantities}
  \label{fig:alt_k1:level:q}
\end{subfigure}%
\begin{subfigure}{.5 \textwidth}
  \centering
  \includegraphics[width=0.95\linewidth]{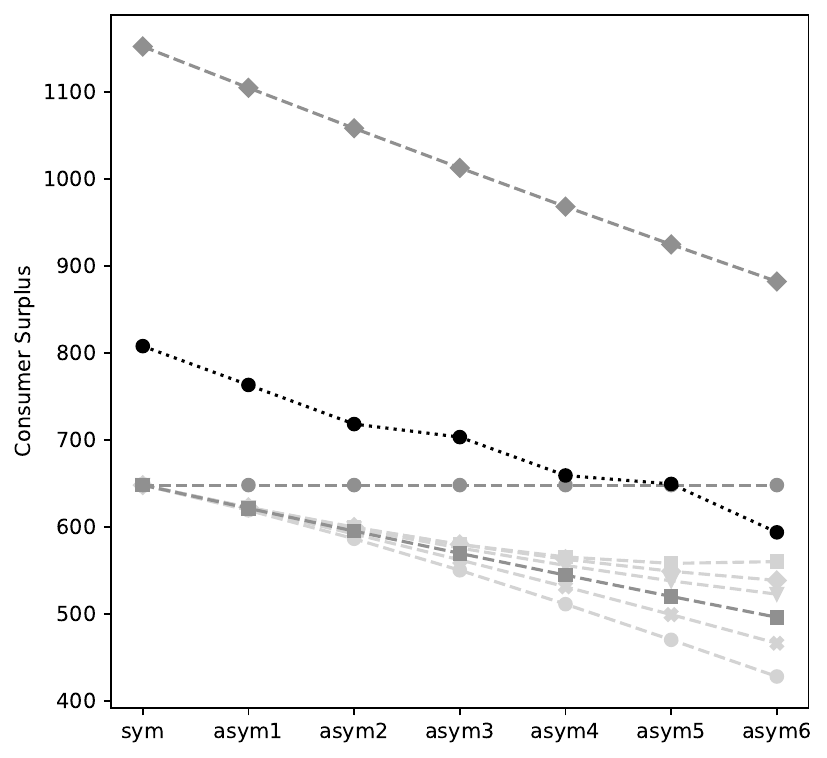}
  \caption{Consumer Surplus}
   \label{fig:alt_k1:level:cs}
\end{subfigure}
\begin{subfigure}{.5 \textwidth}
  \centering
  \includegraphics[width=0.95\linewidth]{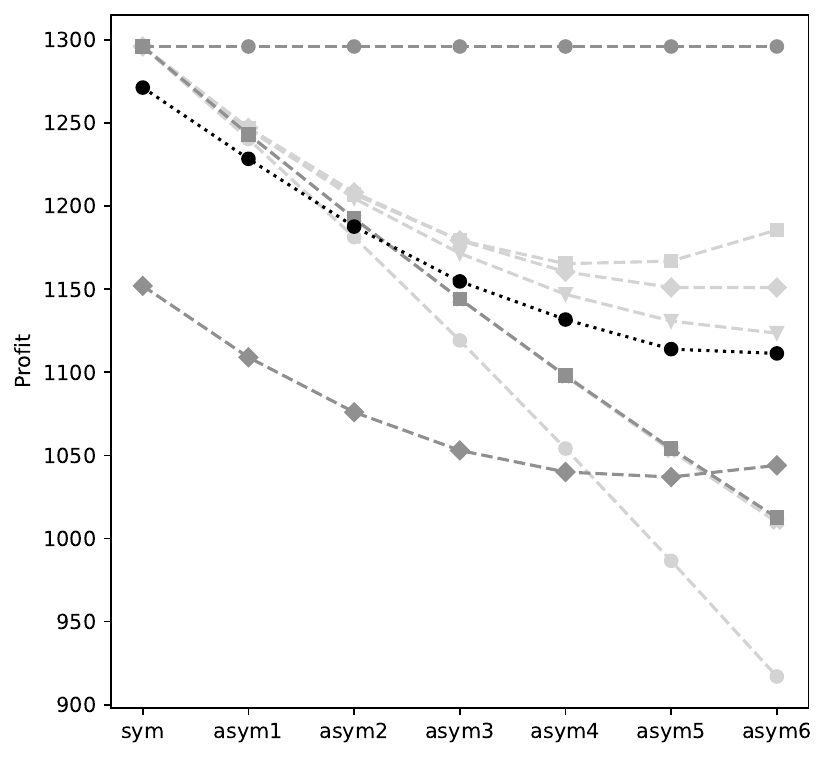}
  \caption{Total Profits}
  \label{fig:alt_k1:level:pi}
\end{subfigure}%
\begin{subfigure}{.5 \textwidth}
  \centering
  \includegraphics[width=0.95\linewidth]{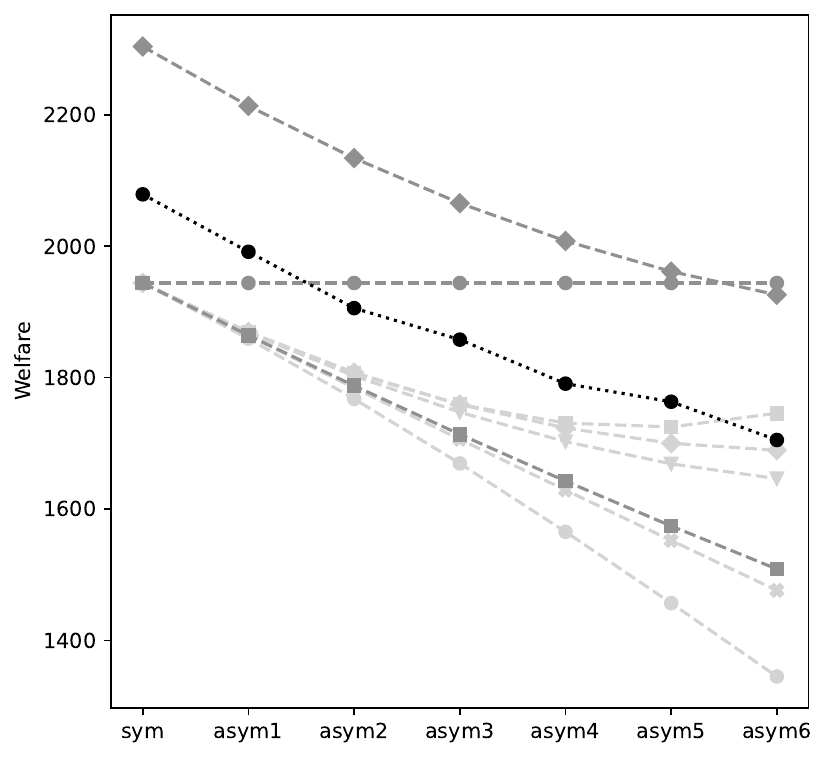}
  \caption{Total Welfare}
\end{subfigure}
\begin{subfigure}{.95 \textwidth}
  \centering
  \includegraphics[width=0.85\linewidth]{img_alt_k1/legend.pdf}
\end{subfigure}
\caption{Simulation results (levels) for alternative different degrees of asymmetry without memory.}
\label{fig:alt_k1:res:1}
\end{center}
\end{figure}

\begin{figure}[H]
\begin{center}

\centering
\includegraphics[width=0.7\linewidth]{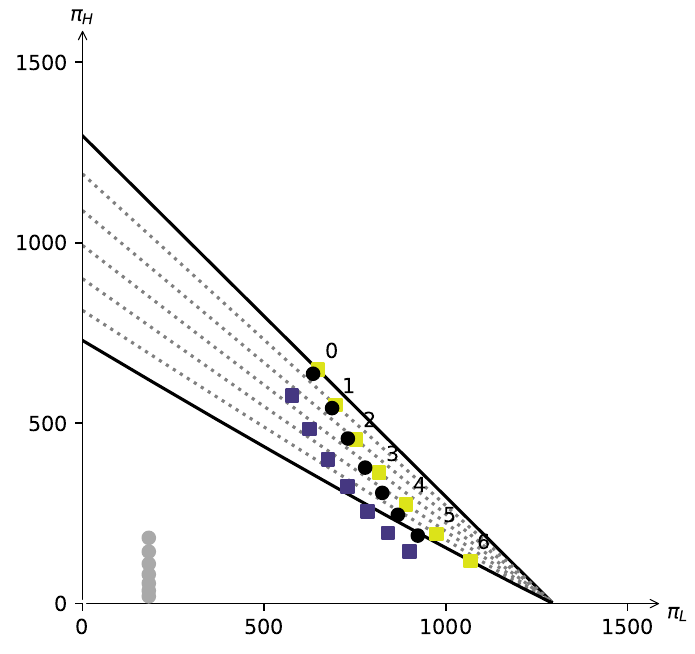}
\caption{Pareto frontier and simulation results for alternative different degrees of asymmetry with (symmetric case: 0).}
\label{fig:pareto:alt_k1}

\end{center}
\end{figure}

\begin{table}[htb!]
\centering
\caption{Average squared distances (alternative cost parameterization)}
\label{tab:alt_k1:average_distances}
\begin{tabular}{lrrrr}
\hline
 & Q & CS & $\Pi$ & W \\
\hline
Nash & 59 & 100,091 & 10,024 & 47,206 \\
Monopoly & 5 & 7,111 & 18,731 & 20,373 \\
Alternating Monopoly & 13 & 16,849 & 2,207 & 23,635 \\
Kalai-Smorodinsky & 16 & 19,550 & 2,310 & 27,943 \\
Equal Split & 20 & 24,063 & 8,868 & 52,103 \\
Equal Relative Gains & 10 & 13,285 & 1,626 & 8,528 \\
Kalai-Smorodinsky (Nash) & 11 & 14,896 & 309 & 10,931 \\
Equal Relative Gains (Nash) & 10 & 13,848 & 823 & 8,750 \\
\hline
\end{tabular}
\end{table}

\newpage
\begin{table}[htb!]
\centering
\caption{Average squared normalized distances (multiplied by 1,000)}
\label{tab:alt_k1:average_distances_normalized}
\begin{tabular}{lrrrr}
\hline

 & Q & CS & $\Pi$ & W \\
\hline
Nash & 0.11 & 0.38 & 0.26 & 0.07 \\
Monopoly & 6.99 & 24.95 & 8.15 & 13.52 \\
Alternating Monopoly & 0.11 & 0.38 & 2.19 & 0.55 \\
Kalai-Smorodinsky & 0.09 & 0.31 & 2.27 & 1.00 \\
Equal Split & 0.64 & 1.89 & 7.07 & 4.38 \\
Equal Relative Gains & 1.11 & 3.61 & 0.34 & 1.25 \\
Kalai-Smorodinsky (Nash) & 0.41 & 1.31 & 0.02 & 0.17 \\
Equal Relative Gains (Nash) & 0.70 & 2.28 & 0.06 & 0.58 \\
\hline
\end{tabular}
\end{table}

\end{document}